\documentclass[aps,prx,
groupedaddress,showpacs,nofootinbib,amssymb,balancelastpage,preprintnumbers]{revtex4}

\usepackage[dvipdfmx]{graphicx}
\usepackage{graphicx,bm,color}

\usepackage{amsmath}
\usepackage{amssymb}
\usepackage{amsfonts}
\usepackage{cases}
\usepackage{cancel}
\usepackage{hyperref}
\usepackage{ulem}
\usepackage{subfigure}
\usepackage{here}
\usepackage{color}
\usepackage{tikz}
\usetikzlibrary{matrix}

\begin{document}

\title{
QCD knows new quarks 
}  

\author{Chuan-Xin Cui}\thanks{{\tt cuicx1618@mails.jlu.edu.cn}}
\affiliation{Center for Theoretical Physics and College of Physics, Jilin University, Changchun, 130012,
China}

\author{Hiroyuki Ishida}\thanks{{\tt ishidah@pu-toyama.ac.jp}}
	\affiliation{Center for Liberal Arts and Sciences, Toyama Prefectural University, Toyama 939-0398, Japan}

\author{Mamiya Kawaguchi}\thanks{{\tt kawaguchi@fudan.edu.cn}} 
      \affiliation{ 
School of Nuclear Science and Technology, University of Chinese Academy of Sciences, Beijing 100049, China
}

\author{Jin-Yang Li}\thanks{{\tt lijy1118@mails.jlu.edu.cn}}
\affiliation{Center for Theoretical Physics and College of Physics, Jilin University, Changchun, 130012,
China}

\author{Shinya Matsuzaki}\thanks{{\tt synya@jlu.edu.cn}}
\affiliation{Center for Theoretical Physics and College of Physics, Jilin University, Changchun, 130012,
China}

\author{Akio Tomiya}\thanks{{\tt akio@yukawa.kyoto-u.ac.jp}}
\affiliation{RIKEN BNL Research center, Brookhaven National Laboratory, Upton, NY, 11973, USA}
\affiliation{Department of Information Technology, International Professional University of Technology, Osaka, 3-3-1, Umeda, Kita-Ku, Osaka, 530-0001, Japan}

\begin{abstract}

We find that a big gap between indicators for the breaking strengths of the global chiral SU(2) and U(1) axial symmetries in QCD of the Standard Model (SM) 
can be interpreted as a new fine-tuning problem. 
This may thus imply calling for a class of Beyond the SM, which turns out to favor  having a new chiral symmetry and the associated massless new quark 
insensitive to the chiral SU(2) symmetry for the lightest up and down quarks, 
so that the fine-tuning is relaxed. 
Our statistical estimate shows that QCD of the SM is 
by more than 300 standard deviations 
off the desired parameter space, which is free from the fine-tuning, and the significance 
will be greater as the lattice measurements on the QCD hadron observables 
get more accurate. 
As one viable candidate, we introduce 
a dark QCD model with massless new quarks,  
which can survive current experimental, cosmological, and astrophysical limits, 
and also leave various phenomenological 
and cosmological consequences, to be probed in the future. 
This is a new indication from QCD, which gives a new avenue to deeper understand QCD, and 
provides a new guideline to consider going beyond the SM.

\end{abstract} 
\maketitle


\section{Introduction}

QCD possesses the intrinsic scale of $ {\cal O}(1)$ GeV, at 
which scale the QCD interaction gets strong enough to form hadrons 
along with the color confinement. 
All the physical quantities in hadron physics should therefore 
arise associated with this order of ${\cal O}(1)$ GeV. 
However, e.g., the mass difference between proton ($p \sim uud$) and neutron ($n \sim udd$) is not of this order:  individual masses, being mainly fed by the isospin-symmetric dynamical quark mass, 
are indeed of ${\cal O}(1)$ GeV, but the mass difference is of ${\cal O}(10^{-3})$ GeV. 
This is thought of as a fine-tuning.

We now know that the tiny mass difference is due to small violation of the isospin symmetry for  
up and down quarks, including the current quark masses $m_u$ and $m_d$ arising from 
the Higgs via the electroweak symmetry breaking and their electromagnetic charge difference.  
The mass difference in fact goes to zero in the symmetric limit,  
and, in this sense, the relaxation of the fine-tuning is guaranteed within QCD, or the SM 
of particle physics alone, without invoking new physics.

The second example is the extraordinary lightness of isotriplet pions composed as 
the bound states of up and down quarks (e.g. $\pi^0 \sim \bar{u}u + \bar{d}d$). 
The pion masses have been observed to be $\sim 140$ MeV, which is 
by one order of magnitude smaller than the 1 GeV scale. 
Then a question is: why is it so smaller? 
This question is also actually trivial and self-resolved within QCD. 
It is 
explained by a small size of explicit violation of the chiral symmetry for 
the up and down quarks. 
In fact, the established chiral perturbation theory~\cite{Gasser:1983yg,Gasser:1984gg} 
shows that due to the chiral symmetry, the pion mass keeps small even against quantum corrections 
arising from the strong coupling scale of ${\cal O}(1)$ GeV, 
due to the chiral symmetry. 
This pion is thus called technical natural~\cite{Wells:2013tta} a la `t Hooft~\cite{tHooft:1979rat}.

Thus the notion of symmetry can make a big gap in scales filled to relax the 
fine-tuning.  

Several types of fine-tunings  
have so far been pointed out, 
which cannot be resolved by the SM of particle physics or cosmology alone. 
All those involve    
an unsatisfactory big cancellation;   
e.g. the gauge hierarchy problem~\cite{Shaposhnikov:2007nj,Nielsen:2012pu,Masina:2012tz,Degrassi:2012ry} 
and the strong CP problem~\cite{Peccei:1977ur,Shifman:1979if,Kim:2008hd}, 
where the former belongs to the category of technical unnaturalness, while the latter does not because the QCD-theta term does not get renormalization in the continuum theory. 
The associated fine-tuned small observables 
have been confirmed: the size of the Higgs mass much smaller than the Planck scale, and   
the yet unobserved electromagnetic dipole moment of neutron, 
respectively. 
This class of fine-tunings  
can generically be related to existence of a hidden new symmetry which 
relaxes the big cancellation, so that 
the theory becomes free from the fine-tuning in the symmetric limit. 
Taking them seriously into account has so far motivated 
people to refine or go beyond the standard theories with such a new hidden symmetry, 
and opened numerous frontiers in research directions along the theoretical particle 
and cosmological physics.

In this paper, we pose a new and nontrivial fine-tuning problem in QCD of the SM, 
and propose a new hidden symmetry which relaxes the fine-tuning.

QCD has been well explored and confirmed, but actually we know less precisely how low-energy QCD and the vacuum depends on quark flavors (described like the Columbia plot), in particular, little understand how the relatively heavy strange quark contributes there. 
This important open issue is thought of as an analogy to the top quark contribution to the electroweak-symmetry broken vacuum in the Higgs potential of the Standard Model, called 
the electroweak-vacuum stability problem. 
What the present work focuses on is 
such a still nontrivial quark-flavor structure of the QCD vacuum, in particular, 
the essential gap between the breaking strengths of the chiral $SU(2)$ symmetry for the light up and down quarks and $U(1)$ axial symmetry.

A finely tuned big gap is found in 
three fundamental quantities of QCD: 
susceptibility functions for 
the chiral $SU(2)_L \times SU(2)_R$ symmetry, 
the $U(1)_A$ axial symmetry, and the topological susceptibility, 
which are essential to characterize 
the vacuum structure of QCD with the lightest three flavors (up, down, and strange quarks).  
The chiral symmetry is operative only for up and down quarks, while the latter two 
are correlated by their axial charges. 
Those susceptibilities are not direct observables in 
terrestrial experiments nor astrophysical observatories, 
in contrast to the existing fine-tuning problems as aforementioned. Those can rather be observed in the lattice QCD, though would not 
have correlation with definite phenomenological observables.

The three susceptibilities are robustly related to each other by the anomalous Ward identities for the chiral $SU(3)_L \times SU(3)_R$ symmetry, 
to hold the form symbolically like 
\begin{align} 
\langle \text{Chiral  SU(2)} \rangle = \langle \text{U(1) Axial} \rangle - \langle \text{Topological} \rangle 
\,, \label{symbolic-trilemma}
\end{align}
where brackets stand for vacuum expectation values.  
(For the precise expressions written in terms of the 
susceptibilities, see Eq.(\ref{WI-def}) in the later section, Sec.~II.)  
$\langle \text{Chiral  SU(2)} \rangle$ and $\langle \text{U(1) Axial} \rangle$ 
are indicators of breaking strengths for the chiral $SU(2)_L \times SU(2)_R$ symmetry and  
$U(1)_A$ axial symmetry
that go to zero, when those symmetries are restored, respectively. 
The new fine-tuning reads presence of 
a big gap in magnitude between 
$\langle \text{Chiral  SU(2)} \rangle$ and $\langle \text{U(1) Axial} \rangle$, because of nonzero and sizable 
$\langle \text{Topological} \rangle $ 
in QCD. 
\textcolor{black}{
 There a drastic cancellation between two independent infrared singularities is observed, which is responsible for existence of the soft pions and required to yield the finite quark condensate. 
 This thus causes the fine-tuning and yields a gigantic gap between 
 the chiral $SU(2)$ and $U(1)$ axial breaking strengths. 
}

To demonstrate the new fine-tuning in a quantitative way, 
one needs to work on QCD in the deep infrared region, which is, however, 
highly nonperturbative because of the strong coupling nature in the low-energy 
scale. 
The best method to compute such nonperturbative dynamics 
is the numerical simulations of QCD on lattices, which have however never 
measured the three susceptibilities in Eq.(\ref{WI-def}) on the same lattice 
setting at the same time.

Instead of the lattice simulation, we can invoke effective models of low-energy QCD, 
on the spirit of Weinberg~\cite{Weinberg:1978kz}, which realize  
the same breaking structure of the chiral and axial symmetries, and so forth, 
as that in the low-energy QCD. 
In this paper, as the low-energy QCD description, we thus adapt a class of 
the Nambu-Jona Lasinio (NJL) model made of only quarks with several 
quarkonic interactions. 
The NJL model has extensively been utilized in the field of 
hadron physics, and so far provided us with lots of qualitative interpretations 
for the low-energy QCD features, associated with the chiral and axial symmetry breaking, 
together with successful phenomenological predictions~\cite{Hatsuda:1994pi}.

We first show statistical good fitness of the model with the lattice simulation data on hadronic 
observables for QCD with three flavors at physical point. 
We then regard the model prediction  
to the three susceptibilities ($\langle \text{Chiral  SU(2)} \rangle$, 
$\langle \text{U(1) Axial} \rangle$, and $\langle \text{Topological} \rangle $) 
as the prediction from full QCD of the SM, with possible theoretical uncertainty. 
We find more than 300 standard deviation for QCD in the standard model away from 
the desired parameter space free from the fine-tuning.


The big gap between ($\langle \text{Chiral  SU(2)} \rangle$ 
and $\langle \text{U(1) Axial} \rangle$ 
vanishes only in the limit where the strange quark mass $m_s$ is sent to zero, 
no matter what nonzero small values of $m_u$ and $m_d$ are taken, as long as the chiral symmetry acts as 
a good symmetry. 
The symmetry, which sends $m_s$ to zero,  
has nothing to do with the chiral and isospin 
symmetries that make QCD fine-tuning free in a view of 
the existing hadron spectra, as noted above. 
More remarkably, the new fine-tuning is present even for 
the two-flavor QCD with quenched strange quark ($m_s \to \infty$), and still survives even when 
the theory approaches the massless two-flavor limit with $m_u, m_d \to 0$.

Observing that the strange quark acts as a spectator for 
the chiral $SU(2)$ symmetry for up and down quarks, 
we deduce that in place of the strange quark, adding a massless new chiral-singlet quark protected by a new symmetry 
makes QCD free from the new fine tuning. 
It is also interesting to note that 
presence of such massless new quarks 
can potentially solve the strong CP problem as well, 
in the same way as the massless up quark solution~\cite{Peccei:1977hh,Peccei:1977ur}.

As one viable candidate to make QCD with any $m_l$ and $m_s$ free from the fine-tuning, 
we address a dark QCD model with massless new quarks.  
The model possesses a new chiral symmetry, 
which protects the new quarks from being massive at the classical level, 
to be broken by quantum anomalies via QCD gluon and dark QCD 
gluon interactions. 
The dark QCD acts like a mirror of QCD, to keep 
the massless new quark contribution to the anomalous Ward-identity as 
in Eq.(\ref{symbolic-trilemma}) below the scale of order of 1 GeV.  
\textcolor{black}{This model is shown to} survive current experimental, astrophysical, and cosmological constraints. 
Several smoking-guns of the presently introduced benchmark model are also discussed.

Several future prospects 
are finally commented in the section describing our conclusion.

\section{Chiral Ward identities and topological susceptibility}

We begin by introducing the key equation showing   
a relation between indicators of the breaking strengths for the chiral $SU(2)_L \times SU(2)_R$ symmetry    
and $U(1)_A$ axial symmetry, together with the topological susceptibility.

We first introduce a set of generic anomalous Ward identities for the three-flavor chiral $SU(3)_L \times SU(3)_R$ symmetry in QCD~\cite{Nicola:2016jlj,Kawaguchi:2020qvg,Cui:2021bqf} 
(see also Appendix~\ref{App:B}):  
\begin{align} 
 \langle \bar uu \rangle +\langle \bar dd \rangle 
 &= - m_l \chi_\pi 
\,, \notag\\  
\langle \bar uu \rangle +\langle \bar dd \rangle
+ 4 \langle \bar s s \rangle
& = 
- \Bigg[[ m_l 
\left(\chi_{P}^{uu}+\chi_{P}^{dd}+2\chi_{P}^{ud}\right)
\notag\\ 
& - 2 (m_s + m_l)\left(\chi_{P}^{us}+\chi_{P}^{ds} \right)
+ 4 m_s \chi_P^{ss} \Bigg] 
\,, \notag\\ 
\langle \bar uu \rangle +\langle \bar dd \rangle
-2  \langle \bar s s \rangle
& = 
- \Bigg[ m_l 
\left(\chi_{P}^{uu}+\chi_{P}^{dd}+2\chi_{P}^{ud}\right)
\notag\\ 
& 
+  (m_l - 2 m_s)\left(\chi_{P}^{us}+\chi_{P}^{ds} \right) 
-  2 m_s \chi_P^{ss} \Bigg] 
		\label{chipi}\,, 
	\end{align} 
	where the isospin symmetric limit $m_u=m_d\equiv m_l$ has been taken,   
	$\chi_p^{uu}, \chi_P^{dd}, \chi_P^{ud}$, $\chi_P^{ss}$, $\chi_P^{us}$, and $\chi_P^{ds}$ are the pseudoscalar susceptibilities and $\chi_\pi$ is the pion susceptibility, that are defined as 
\begin{align} 
		\chi_P^{f_1f_2} &=\int  d^4 x \langle (\bar q_{f_1}(0) i \gamma _5 q_{f_1}(0))(\bar q_{f_2}(x) i \gamma _5 q_{f_2}(x))\rangle\,, 
\notag\\ 
& {\rm for} \quad {q_{_{f_{1,2}}} = u,d,s}\,, \notag\\ 
	\chi_{\pi}
	&= \int  d^4 x 
	\Bigg[ 
	\langle ( \bar u(0) i \gamma_5  u(0))( \bar u(x) i\gamma_5 u(x))\rangle_{\rm conn}
\notag\\ 
& 
+ \langle ( \bar d(0)  i\gamma_5 d(0))( \bar d(x) i\gamma_5 d(x))\rangle_{\rm conn}
\Bigg] 
\,, 		\label{psesus}
	\end{align} 
with $\langle \cdot \cdot \cdot \rangle_{\rm conn}$ being the connected part of the correlation function. 

Second, we introduce the topological susceptibility $\chi_{\rm top}$, which is related to the $\theta$ vacuum configuration of QCD. 
It is defined as the curvature of the $\theta$-dependent vacuum energy $V(\theta)$ in QCD at  $\theta=0$:
	\begin{equation}
		\chi_{\rm top}=-\int d^{4}x \frac{\delta^2 V(\theta)}{\delta \theta(x)\delta \theta(0) }\Bigg{|}_{\theta =0}
		\,. \label{chitop:def}
	\end{equation}
Performing the $U(1)_A$ rotation for quark fields 
together with the flavor-singlet condition \cite{Baluni:1978rf,Kim:1986ax}, one can transfer the $\theta$ dependence coupled to the topological gluon configurations, via the axial anomaly, into current quark mass terms. 
Thus $\chi_{\rm top}$ goes like \cite{Kawaguchi:2020qvg}
	\begin{align}
		\chi_{\rm top} 
		&=
		 \bar{m}^2 \Bigg[ 
		\frac{\langle \bar {u}u \rangle}{m_l}  
		+\frac{\langle \bar {d}d \rangle}{m_l} 
		+\frac{\langle \bar {s}s \rangle}{m_s} 
\notag\\
& 
+ \chi_{P}^{uu}+\chi_{P}^{dd}
		+\chi_P^{ss} 
		+ 2 \chi_P^{ud} 
		+2\chi_{P}^{us}+
		2\chi_{P}^{ds} 
        \Bigg] 
		\label{chitop}
	\notag \\ 
	& = \frac{1}{4} \left[ 
	m_l\left(\langle \bar{u} u \rangle +\langle \bar{d} d \rangle  \right)
	+ m_l^2\left(\chi_{P}^{uu}+\chi_{P}^{dd}
	+2\chi_{P}^{ud}
	\right)
	\right] 
\notag\\ 
&	= m_s \langle \bar{s}s \rangle + m_s^2 \chi_P^{ss} 
	\,, 
	\end{align}
\noindent 
		where $\bar m \equiv \left(\frac{1}{m_u}+\frac{1}{m_d}+\frac{1}{m_s} \right)^{-1}$.   
Throughout the present paper, 
we take the signs of quark condensates and quark masses to be negative and positive, respectively, so that 
$\chi_{\rm top} <0$ (hence $\langle {\rm Topological} \rangle$ in Eq.(\ref{symbolic-trilemma}) contributes as a negative term).  
Note that $\chi_{\rm top} \to 0$, when either of quarks becomes massless ($m_l$ or $m_s$ $\to 0$), reflecting the flavor-singlet nature of the QCD vacuum.

By combining Ward identities in Eq.(\ref{chipi}), 
$\chi_{\rm top}$ in Eq.(\ref{chitop:def}) is expressed to be  
	\begin{equation}
	\begin{aligned}
		\chi_{\rm top} &= 
		 \frac{1}{2} m_l m_s
		\left(\chi_{P}^{us}+\chi_{P}^{ds} \right)
		=\frac{1}{4}  m_l^2 ( 
		\chi_\eta 
		-\chi_{\pi})
		\,,  
	\end{aligned}
	\label{disc}
	\end{equation} 
	where $\chi_\eta$ is the $\eta$ meson susceptibility, which is defined as 
\begin{align}
	\chi_{\eta} &=\int d^4 x 
	\Bigg[ 
	\langle (\bar u(0) i \gamma _5  u(0))
	(\bar u(x) i\gamma _5 u(x))\rangle 
\notag\\
	& +
	\langle (\bar d(0) i\gamma _5  d(0))
	(\bar d(x) i\gamma _5 d(x))\rangle
\notag\\ 
&	+
	2 
	\langle (\bar u(0) i\gamma _5  u(0))
	(\bar d(x) i\gamma _5 d(x))\rangle
	\Bigg] 
	\notag\\ 
	& = \chi_{P}^{uu}+\chi_{P}^{dd}+2\chi_{P}^{ud}
	\,. \label{chi-eta}
\end{align} 
Equation (\ref{disc}) can be written as  
\begin{align} 
(\chi_\eta - \chi_\delta)  =   (\chi_\pi - \chi_\delta) + \frac{4}{m_l^2} \chi_{\rm top}  
 \,, \label{WI-def}
\end{align}
where $\chi_\delta$ is the susceptibility for the 
$\delta$ meson channel (which is $a_0$ meson in terms of the Particle Data Group identification), defined in the same way as $\chi_\pi$ in Eq.(\ref{chipi}) 
with the factors of $(i \gamma_5)$ replaced with identity $1$. 
Renaming $(\chi_\eta - \chi_\delta)$ and $(\chi_\pi - \chi_\delta)$ as 
\begin{align} 
 \chi_{\rm chiral} 
 & \equiv \chi_\eta - \chi_\delta \,, \notag\\ 
  \chi_{\rm axial} 
 & \equiv \chi_\pi - \chi_\delta
\,,  \label{chiral-axial}
\end{align}
which corresponds respectively to $\langle \textrm{Chiral SU(2)} \rangle$ and $\langle \textrm{U(1) axial} \rangle$ 
in Eq.(\ref{symbolic-trilemma}), in the Introduction.

In deriving Eq.(\ref{WI-def}) one could choose another scalar susceptibility $\chi_{\sigma}$, 
which makes the chiral $SU(2)$ and axial 
partners for $\chi_{\pi}$ and $\chi_\eta$, respectively. [The definition of $\chi_\sigma$ is the same as $\chi_\eta$ with the $(i \gamma_5)$ factors replaced by the identity $(1)$.] 
Then Eq.(\ref{WI-def}) would be replaced by 
$ \chi_{\rm chiral}' = \chi_{\rm axial}' + \frac{4 \chi_{\rm top}}{m_l^2}$, 
where $\chi_{\rm chiral}' \equiv 
\chi_{\sigma} - \chi_\pi$ and $\chi_{\rm axial}' \equiv \chi_\sigma - \chi_\eta$. 
Even if this alternative identification is taken, the present proposal of the new fine-tuning problem still holds: 
a big gap between $\chi_{\rm axial}'$ and the $\chi_{\rm chiral}'$ is relaxed by a symmetry sending   
$m_s \to 0$ to make $\chi_{\rm top}/m_l^2$ vanishing.

\begin{figure} 
\begin{center}
\begin{tikzpicture}
  \matrix (m) [matrix of math nodes,row sep=4em,column sep=5em,minimum width=3em]
 {
\chi_{\pi} &  \chi_{\delta} & \chi_{\eta}\\
};
  \path[-stealth]
     (m-1-1) edge  node [above] {U(1) Axial} (m-1-2)
    (m-1-2) edge node [midway,left] { } (m-1-1)
    (m-1-3) edge  node [above] {Chiral SU(2)} (m-1-2)
    (m-1-2) edge node [midway,left] { } (m-1-3)
;
\end{tikzpicture}
\end{center} 
\caption{A schematic cartoon on 
the chiral and axial transformations for susceptibilities} 
\label{chiral-axial-fig} 
\end{figure}

\section{New fine-tuning problem and hidden symmetry}

Now we discuss a new fine-tuning problem based on the key Eq.(\ref{WI-def}).

First of all, consider In the case with small enough $m_l$ and finite strange quark mass $m_s$ 
($m_s \gg m_l \to 0$), as 
in QCD of the SM at physical point, 
the topological susceptibility $\chi_{\rm top}$ 
can approximately be evaluated as~\footnote{
Throughout the present Letter, 
we take the signs of quark condensates and quark masses to be negative and positive, respectively, so that 
$\chi_{\rm top} <0$. 
} 
\begin{align}
 \chi_{\rm top} \Bigg|_{m_l \ll m_s} \sim 
 \bigg{(} 
 \frac{\langle\bar{u}u\rangle}{m_l}
 +\frac{\langle\bar{d}d\rangle}{m_l}
		+\frac{\langle \bar {s}s \rangle}{m_s} \bigg{)}\bar{m}^2 
		\,, \label{chitop-small-ml}     
\end{align} 
where $\bar{m} = (2/m_l + 1/m_s)^{-1}$. 
Noting that 
$\langle\bar{u} u\rangle$, 
$\langle\bar d d\rangle$,  
and $\langle\bar{s}s\rangle$ keep nonzero even 
when $m_l=0$, because of the dynamical generation of quark condensates in QCD at the scale of ${\cal O}(1)$ GeV, 
we find that 
for small $m_l \ll m_s$, the $\langle \bar{u}u \rangle$ 
and $\langle \bar{d}d \rangle$ terms are dominant in Eq.(\ref{chitop-small-ml}), so that 
the $\chi_{\rm top}$ term in Eq.(\ref{WI-def}) is well approximated as 
\begin{align} 
 \frac{\chi_{\rm top}}{m_l^2} \Bigg|_{m_l \ll m_s} 
 & \sim 
 \left(
  \frac{\langle\bar{u}u\rangle}{m_l}
 +\frac{\langle\bar{d}d\rangle}{m_l}
 \right)
 \frac{\bar{m}^2}{m_l^2}  
\notag\\ 
& \sim  - \frac{[{\cal O}(1) \, {\rm GeV}]^3}{4 m_l}
		\,,  \label{chitop-ms-ml-scaling}
\end{align}
with the minus sign of the quark-condensate value taken into account. 
Thus the size of the $\chi_{\rm top}$ term gets larger than $[{\cal O}(1) \, {\rm GeV}]^2$ 
(with minus sign). 
Note also that 
$\chi_{\rm chiral} > 0$, $\chi_{\rm axial} >0$, and $\chi_{\rm chiral} 
< \chi_{\rm axial}$ due to the 
measured meson spectroscopy (for more details, see Appendix~\ref{App:A}). 
Therefore, in the case with small $m_l$ and finite $m_s$, 
we meet a {\it big destructive cancellation} in Eq.(\ref{WI-def}) between 
$\chi_{\rm axial}$ and the $\chi_{\rm top}$ term, 
both of which are on the order bigger than $[{\cal O}(1) \, {\rm GeV}]^2$, to have a highly suppressed 
$\chi_{\rm chiral}$:  Equation (\ref{WI-def}) looks like 
$[{\cal O}(10 \, {\rm GeV})]^2_{\chi_{\rm axial}} - [{\cal O}(10 \, {\rm GeV})]^2_{\chi_{\rm top}} = \chi_{\rm chiral} \ll [{\cal O}(1\, {\rm GeV})]^2$ for $m_l \lesssim $ MeV. 
This can be interpreted as a fine-tuning unless some symmetry is present to 
explain the extraordinary small $\chi_{\rm chiral}$ which can relax 
the big subtraction, as elaborated in Introduction. 
Note, however, that 
even the conventional chiral $SU(2)$ symmetry ($m_l \to 0$) makes the accidental big cancellation more serious. 
As it will turn out later, 
QCD of the SM with  
light up and down quarks, and relatively heavier strange quark actually 
suffers from this kind of big subtraction.

This conclusion is unambiguous and would not be altered even if 
the Ward identity in Eq.(\ref{WI-def}) were subtracted by 
another scalar susceptibility $\chi_\sigma$ (which is defined 
as in the same manner as $\chi_\eta$ in Eq.(\ref{chi-eta}) with the 
$(i \gamma_5)$ factors replaced by identity, and 
forms the chiral partner of $\chi_\pi$ and the axial partner of 
$\chi_\eta$.)

\textcolor{black}{The existence of the fine-tuning is due to the accidental cancellation between two individual infrared singularities responsible for 
the soft pions in QCD:  
as discussed in the literature~\cite{Cui:2021bqf}, $\chi_{\rm axial} \sim 1/m_{\pi}^2 \sim 1/m_l$, $\chi_{\rm chiral} \sim$ constant, and $\chi_{\rm top}/m_l^2 \sim 1/m_l$ 
(see also Eq.(\ref{chitop-ms-ml-scaling})) for $m_l \ll m_s$ and $m_l \to 0$, 
hence in this limit Eq.(\ref{WI-def}) looks like finite = $\infty - \infty$. 
This observation may imply that the chiral limit, on the base of which QCD can be 
expanded in a way of the chiral perturbation, hence is widely accepted and well established, is faced with an accidental fine-tuning. 
Thus the proposed fine-tuning is completely separated from 
the already existing fine-tuning, e.g., on  
the tiny mass difference between proton and neutron. }

Going away from QCD of the SM, 
we shall consider a counter limit where $m_s \to 0$ 
with keeping finite $m_l$. 
In Eq.(\ref{WI-def}) the $\chi_{\rm top}$ term then goes vanishing as $m_s \to 0$ [See also Eq.(\ref{chitop-small-ml})] reflecting 
the flavor-singlet nature~\cite{Cui:2021bqf}, 
so that the indicators for the breaking strengths of the chiral and axial symmetries 
become identically 
equal each other: 
\begin{align} 
( \chi_{\rm chiral} - \chi_{\rm axial}) = \frac{4 \chi_{\rm top}}{m_l^2}\to 0 
\,,  \qquad {\rm as} \, \quad  m_s \to 0.   
\label{tech-natur-def}
\end{align}
In this case the $\chi_{\rm top}$ term $(\chi_{\rm top}/m_l^2)$ 
is adjusted to zero  by a big destructive subtraction, i.e., 
a fine-tuning between 
$\chi_{\rm chiral}$ and $\chi_{\rm axial}$. 
However, this fine-tuning can be gone  
in the limit $m_s \to 0$ which makes  
$(\chi_{\rm top}/m_l^2)$ sent to zero, in contrast to QCD of the SM argued above, 
though the case with $m_l \gg m_s \to 0$, where $\chi_{\rm chiral} \sim \chi_{\rm axial}$, 
is unrealistic.

Therefore, 
the presently addressed fine-tuning has nothing essentially to do with the smallness of 
up and down quarks, i.e., the existing chiral symmetry. 
Hence it is completely separated from 
the already existing fine-tuning, e.g., on  
the tiny mass difference between proton and neutron.

Note that the strange quark currently acts as a spectator for the chiral 
$SU(2)$ symmetry, being singlet. 
Hence introduction of a new massless quark, protected by own chiral symmetry i.e., {\it hidden new symmetry},  should 
play the same role as the strange quark to  
solve the fine-tuning problem, with keeping massive enough strange quark in accordance with the observation. 
We will later introduce an explicit and phenomenologically viable 
model having massless new quarks ($\chi$) with a new chiral symmetry, 
which makes real-life QCD free from the fine-tuning: 
\begin{align} 
( \chi_{\rm chiral} - \chi_{\rm axial}) \to 0 
\,,  \qquad {\rm as} \, \qquad  m_\chi \to 0\,,  
\label{tech-natural-chi} 
\end{align}
with $m_l$ and $m_s$ at physical point.

In the next section we will explicitly demonstrate the new fine-tuning  
of QCD in a more quantitative way.

\section{Quantifying the fine-tuning}

We define a ratio~\cite{Cui:2021bqf} 
\begin{align} 
R \equiv  \frac{\chi_{\rm chiral}}{\chi_{\rm axial}} 
 \,, 
 \label{R-def}
\end{align}
which also reads $ R = 1 - \frac{4 \chi_{\rm top}/m_l^2}{\chi_{\rm axial}} 
= \frac{1}{1 + \frac{4 \chi_{\rm top}/m_l^2}{\chi_{\rm chiral}} }$ via the Ward identity in Eq.(\ref{WI-def}). 
Thus the deviation from $R=1$ dictates a fine-tuning, 
hence $R$ serves as  the estimator of the fine-tuning.

To compute the estimator $R$, 
one needs to work on QCD in the deep infrared region, which is 
highly nonperturbative because of the strong coupling nature in the low-energy 
scale. 
The best method to compute such nonperturbative dynamics 
is the numerical simulations of QCD on the  lattice. 
However, the lattice simulations 
have never measured the susceptibilities at vacuum with 
varying $m_s$~\footnote{
This is mainly because firstly it has not been well motivated, and moreover, costs of lattice calculations 
for small mass are proportional to $1/m$, where $m$ is the lightest quark mass.
Simulations for light strange quarks can be performed using the same technology as in \cite{Dini:2021hug}, and employing similar calculations in \cite{HotQCD:2012vvd, Bhattacharya:2014ara} with light $m_s$. 
}.

Instead of the lattice simulation, 
in the spirit of Weinberg~\cite{Weinberg:1978kz} 
we can invoke effective models of low-energy QCD, which realize  
the same breaking structure of the chiral and axial symmetries, and so forth, 
as that in the low-energy QCD. 
In this paper, as the low-energy QCD description, we thus adapt a class of 
the Nambu-Jona Lasinio (NJL) model made of only quarks with several 
quarkonic interactions. 
The NJL model has extensively been utilized in the field of 
hadron physics, and so far provided us with lots of qualitative interpretations 
for the low-energy QCD features, associated with the chiral and axial symmetry breaking, 
together with successful phenomenological predictions~\cite{Hatsuda:1994pi}.

\section{Best-fit model estimate}

We employ an NJL model with three flavors, which  
takes the form (for a review, see \cite{Hatsuda:1994pi}):
	\begin{align}
		\mathcal{L}& =\bar{q}(i\gamma_{\mu}\partial^{\mu} -{\cal M} )q+\mathcal{L}_{4f}+\mathcal{L}_{\rm KMT} \,, \notag \\
		\mathcal{L}_{4f}&=\frac{G_S}{2}\sum^{8}_{a=0}\lbrack(\bar{q}\lambda^a q)^2+(\bar{q}i \gamma_5 \lambda^a q )^2\rbrack \,, \notag \\
		\mathcal{L}_{\rm KMT}&=G_D\lbrack \mathop{\rm det}\limits_{i,j}\bar{q_i}(1+\gamma_5 )q_j+ {\rm h.c.} \rbrack 
		\,, \label{Lag:NJL}
	\end{align}
	where the quark field $q$ is represented as the triplet of $SU(3)$ group in the flavor space, $q=(u,d,s)^T$, and $\lambda^a$ ($a=0, 1, \cdots, 8$) are the Gell-Mann matrices with $\lambda^0= \sqrt{2/3}\cdot 1_{3 \times 3}$. The determinant in ${\cal L}_{\rm KMT}$ acts on the flavor indices, and ${\cal M}={\rm diag}\{ m_l, m_l, m_s\}$. 
	
$\mathcal{L}_{4f}$ is the standard-scalar four-fermion interaction term with the coupling strength $G_S$. This is the most minimal interaction term involving the smallest number of quark fields for Lorentz scalar and  pseudoscalar channels, which could be generated at low-energy  QCD via the gluon exchange. The $\mathcal{L}_{4f}$ is $U(3)_L\times U(3)_R$ invariant under the chiral transformation: $q \to U \cdot q$ with $U= \exp[ - i \gamma_5 \sum_{a=0}^8  (\lambda^a/2) \theta^a ]$ and the chiral phases $\theta^a$. 
	The mass term in ${\cal L}$ explicitly breaks $U(3)_L\times U(3)_R$ symmetry. 
	The determinant term $\mathcal{L}_{\rm KMT}$ is called  the Kobayashi- Maskawa-‘t Hooft \cite{Kobayashi:1970ji,Kobayashi:1971qz,tHooft:1976rip,tHooft:1976snw} term, which is a six-point interaction induced from the QCD instanton configuration coupled to quarks, with the effective coupling constant $G_D$. 
	This interaction gives rise to the mixing between different flavors and also uplift the $\eta^{\prime}$ mass to be no longer a Nambu-Goldstone boson. The KMT term preserves $SU(3)_L\times SU(3)_R$ invariance (associated with the chiral phases labeled as $a=1, \cdots, 8$), but breaks the $U(1)_A$ (corresponding to $a=0$) symmetry. 

The approximate chiral $SU(3)_L \times SU(3)_R$ symmetry is spontaneously broken down 
to the vectorial symmetry $SU(3)_V$,  
when the couplings $G_s$ and/or $G_D$ get strong enough, by nonperturbatively developing 
nonzero quark condensates $\langle \bar{q}q \rangle \neq 0$, to be consistent with 
the underlying QCD feature. 
The present NJL model monitors the spontaneous breakdown by the large $N_c$ expansion, 
where $N_c$ stands for the number of QCD colors.

    The NJL model itself is a (perturbatively) nonrenormalizable field theory because ${\cal L}_{4f}$ and ${\cal L}_{\rm KMT}$ describe the higher dimensional 
    interactions with mass dimension greater than four.  
    Therefore, a momentum cutoff $\Lambda$ needs to be introduced to make the NJL model regularized.


There are five model parameters that need to be fixed: the light quark mass $m_l$, the strange quark mass $m_s$, the coupling constants $G_S$ and $G_D$, and the (three-) momentum cutoff $\Lambda$. 
Since the present NJL model does not incorporate the isospin breaking as well as 
radiative electromagnetic 
and weak interactions, it would not be suitable to input experimental 
values of QCD observables that implicitly include all those corrections. 
We thus use as inputs observables in lattice QCD with 2 + 1 flavors 
in the isospin symmetric limit at the physical point 
available from the literature~\cite{Aoki:2013ldr,Bali:2021qem}, 
which are exclusive for the gauge interactions external to QCD. 
We apply the least-$\chi^2$ test to fix the parameters 
by using five representative observables as in Table~\ref{chi2-fit:tab}. 
The resultant values of the best-fit model parameters are given in Table~\ref{best-fit-para:tab}. 
The least $\chi^2$ test shows good agreement with the lattice data 
within the 1$\sigma$ uncertainties. 
The best-fit NJL model predicts 
the susceptibilities relevant to $R$ 
in Eq.(\ref{R-def}) of the main text 
as 
$\chi_{\rm chiral} = (2.2784 \pm 0.0026)\times 10^{5} $, 
$\chi_{\rm axial} =  (4.8597 \pm 0.0077)\times 10^{6}$, 
and 
$\chi_{\rm top}/m_l^2 = (-1.158 \pm 0.064)\times 10^{6}$, 
in unit of ${\rm MeV}^2$.


With the best fit parameters in Table~\ref{best-fit-para:tab}, 
the present NJL model predicts $\chi_{\rm top}= (0.025 \pm 0.002 )\, /{\rm fm}^{4}$.  
For this $\chi_{\rm top}$, 
comparison with the results from the lattice QCD simulations with 2 + 1 flavors is 
available, which are  
$\chi_{\rm top}=0.019(9)/{\rm fm}^4$~\cite{Bonati:2015vqz}, and $\chi_{\rm top}=0.0245(24)_{\rm stat}(03)_{\rm flow}(12)_{\rm cont}/{\rm fm}^4$~\cite{Borsanyi:2016ksw}. 
Here, for the latter the first error is statistical, the second one comes from the systematic error, and the third one arises due to changing the upper limit of the lattice spacing range in the fit. 
Although their central values do not agree each other, 
we may conservatively say that the 
difference between them is interpreted as a systematic error from the individual lattice QCD calculation. 
Thus the present NJL model is, in that sense, in good agreement with the lattice QCD results on $\chi_{\rm top}$.  
This supports reliability for the present model to estimate $R$ as 
the QCD prediction, as well as the good fitness of the model with 
hadronic observable data on the lattice QCD.


\begin{table}[t] 
\begin{tabular}{|c|c|c|} 
\hline 
\hspace{30pt} observable \hspace{30pt}  &
\hspace{30pt} lattice data \hspace{30pt}  & 
\hspace{30pt} best-fit value\hspace{30pt}  \\ 
\hline \hline 
$f_\pi$ [MeV] & $ 92.07 \pm 0.99$~\cite{Aoki:2013ldr} & $91.93 \pm 1.35$ \\ 
\hline 
$m_\pi$ [MeV] &$138 \pm 0.3$~\cite{Aoki:2013ldr} & $137.95 \pm 1.95$ \\ 
\hline 
$m_K$ [MeV] & $ 494.2 \pm 0.4$~\cite{Aoki:2013ldr}& $493.50 \pm 8.09$
\\ 
\hline 
$m_\eta$ [MeV] & $554.7 \pm 9.2$~\cite{Bali:2021qem} & $503.82 \pm 15.37$ \\ 
\hline 
$m_{\eta'}$ [MeV] & $930 \pm 21$~\cite{Bali:2021qem} & $978.77 \pm 61.59$ \\ 
\hline 
\end{tabular}
\caption{The result on the least $\chi^2$ statistical test of the present NJL model derived by 
fitting to the lattice QCD data with 
2 + 1 flavors in the isospin symmetric limit at the physical point~\cite{Aoki:2013ldr,Bali:2021qem}. 
The details of the global fit of the NJL model including other available lattice data deserve in another publication.
} 
\label{chi2-fit:tab}
\end{table}

\begin{table}[t] 
\begin{tabular}{|c|c|} 
\hline 
\hspace{50pt} model parameter \hspace{50pt}  & 
\hspace{50pt} best-fit value \hspace{50pt}  \\ 
\hline \hline 
$m_l$ [MeV] &  $5.75 \pm 0.05$\\ 
\hline 
$m_s$ [MeV] &  $130.69 \pm 0.98$\\ 
\hline 
$G_S$ [MeV$^{-2}$] & $(14.34 \pm 0.41) \times 10^{-6}$ \\ 
\hline 
$G_D$ [MeV$^{-5}$] & $(- 18.17 \pm 1.35)\times 10^{-14}$  \\ 
\hline 
$\Lambda$ [MeV] &  $569.4 \pm 16.5$\\ 
\hline 
\end{tabular}
\caption{The best-fit values of the model parameters.}
\label{best-fit-para:tab} 
\end{table}

\begin{figure}[t]
  \begin{center}
   \includegraphics[width=11cm]{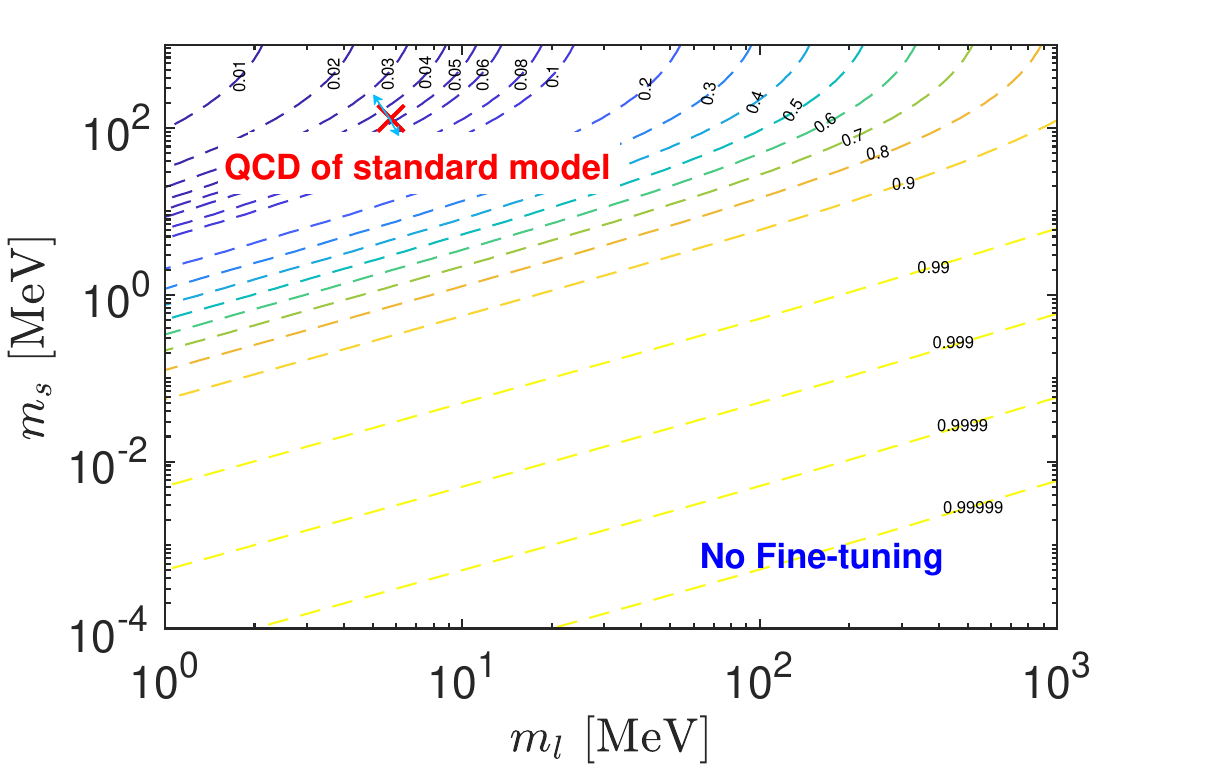}
  \end{center}   
\caption{
This plot visualizes that 
QCD of the SM
is disfavored 
in terms of 
the presently addressed fine-tuning problem.  
The estimated numbers of the estimator $R$ are 
displayed in the $(m_l, m_s)$ plane, 
where the cross mark “$\times$”, labelled as ``QCD of standard model",  
has been plotted by using the best-fit model parameters in 
Table~\ref{best-fit-para:tab}. 
A possible theoretical uncertainty of 30\% for the present NJL model to match full QCD of the 
SM has also been reflected there 
(see also the text), which is drawn 
by light-blue arrows. 
The size of deviations is maximally about 340$\sigma$, and will be at least over 
66$\sigma$ even when 
the theoretical uncertainty of 30\% is considered. 
}  
\label{Ratio-mdep}
\end{figure}

We thus compute the estimator $R$ 
at the best-fit point including the errors associated with the lattice data, 
and find 
\begin{align} 
R = 0.0469 \pm 0.0028 
\, . \label{R-best}
\end{align}
This clarifies that 
{\it QCD at physical point is 
by about 340 standard deviation 
off the desired theory with $R=1$ 
free from the fine-tuning!} 
This is due to too large $m_s$, as noted above.

We may take into account a possible 
theoretical uncertainty of about 30\%,  
which could arise from the leading order approximation 
in the $1/N_c$ expansion, on that 
the present NJL-model prediction is based. 
Currently disregarded corrections, associated with the isospin breaking, 
electromagnetic, and electroweak interactions, 
would also be small enough to be covered by the 30\% uncertainty. 
Therefore, the estimated value of $R$ in Eq.(\ref{R-best}) with the  
theoretical uncertainty of 30\% would be the one corresponding to the prediction of 
the SM. 
Combining this 30\% (``theor.") with the error in Eq.(\ref{R-best}) associated with 
the uncertainties of inputs from the lattice data (``lat."),  
we would then have 
$R = 0.0469 \pm (0.0028)_{\rm lat.} \pm (0.0141)_{\rm theor.}$. 
It is still about 66 standard deviations.

To make this disfavor visualized,  
varying the value of $m_s$ and $m_l$, 
with other model parameters fixed at the best-fit values, 
we plot contours of the estimator $R$ on the $(m_l ,m_s)$ plane, which is displayed in  Figure~\ref{Ratio-mdep}.

The value of $R$ tends to saturate to be $\simeq 0.02$, 
even in the massive two-flavor limit 
with $m_s \to \infty$ and $m_l = 5.75$ MeV. 
This trend is exactly what we have suspected from the $m_s$ scaling of 
the $\chi_{\rm top}$ term in Eq.(\ref{chitop-ms-ml-scaling}). 
With $m_s$ fixed, say, to the physical point, 
$R$ tends to get close to 1 as 
$m_l$ becomes larger, and actually reaches 1 before 
the decoupling limit of 
the up and down quarks, as clarified in the literature~\cite{Cui:2021bqf}.

The deviation in Eq.(\ref{R-best}) can be interpreted as an 
indication of violation of a new symmetry as conjectured in Eq.(\ref{tech-natural-chi}), which the SM does not possess. 
The significance of this new symmetry is subject to 
the accuracy in the current lattice simulation reflected in the size of the error of $R$, 
which is as low as 10\%.  
This significance may therefore be compared to the significance 
for the discovery of the small isospin breaking 
in the $W$ and $Z$ boson masses observed at UA1 and UA2 experiments \cite{UA1:1983crd,UA2:1983tsx,UA1:1983mne,UA2:1983mlz}, which was the same 10\% level in accuracy 
at the final stage of the discovery era (with data taking til 1985)~\cite{UA1:1988rck,UA2:1987qqc}.  
The estimator $R$ can be defined also for the $W$ and $Z$ masses as 
$R_{WZ} \equiv \frac{m_W}{m_Z}$, such that the mass difference is written as  
$\Delta m_{WZ} = m_Z - m_W = m_Z (1 - R_{WZ})$, which actually forms a big cancellation 
structure, hence could be thought of as a fine-tuning, in the same way as in $\chi_{\rm axial}$ and $\chi_{\rm top}/m_l^2$ with $R$ in  Eq.(\ref{R-def}).  
The final UA2 result reads~\cite{UA2:1987qqc} 
$R_{\rm WZ}^{\rm 1982-1985} = 0.876 \pm 0.026$, so it is about 4.8$\sigma$ deviation 
from the isospin symmetric limit $R_{\rm WZ}=1$. 
This is, however, of course trivial and can be explained 
by the isospin breaking (related to the so-called custodial symmetry) in the SM, due to the hypercharge gauge interaction and the presence of isospin breaking in the quark masses. 
Compared to this, the new indication from QCD in Eq.(\ref{R-best}) 
is by about one order of magnitude more significant.

Current precision measurements on $m_W$ and $m_Z$ gives $R_{WZ}^{\rm 2022} = 0.88147 \pm 0.00013$~\cite{Zyla:2020zbs}, which corresponds to 890$\sigma$ deviations from 
the isospin symmetric limit. 
Similarly the error of $R$ in Eq.(\ref{R-best}) is also expected to become 
smaller as the precision in the lattice simulations gets higher in the 
future, hence the significance of the violation of a new symmetry will be enlarged to be as big as the current one for the isospin breaking.

This prospected significance might also become comparable with 
the current significance of the isospin breaking in the proton and neutron mass 
difference with $R_{pn} \equiv \frac{m_p}{m_n} = 1 - \frac{\Delta m_{pn}}{m_n} 
= \frac{1}{1  \frac{\Delta m_{pn}}{m_p}}$, which is read as 
$R_{pn}^{2022} = 0.998623477(316)$~\cite{Zyla:2020zbs}, leading to $\sim 4 \times 10^4\sigma$.

The gauge hierarchy problem caused by the big destructive loop correction to 
the Higgs mass $m_h$ yields $R_H = \frac{m_h^2(m_h)}{m_h^2(M_p)} \approx 
1 -  \frac{3}{8 \pi^2} \frac{M_p^2 - m_h^2}{m_h^2(M_p)}  =  
\frac{1}{1 + \frac{3}{8 \pi^2} \frac{M_p^2 - m_h^2}{m_h^2}}$, where 
$M_p$ denotes the Planck scale $\sim 10^{18}$ GeV, $m_h(m_h) \sim 125$ GeV, 
and we have simply taken into account only the top loop with the top Yukawa coupling $\sim 1$.  
This $R_H$ is estimated to be $\sim 6.3 \times 10^{-32}$. 
A similar estimate can be done also for the strong CP problem, which would yield  $R_{\theta} 
= \frac{\theta_{\rm QCD}}{\theta_{\rm EW}}< 10^{-10}$, 
where $\theta_{\rm QCD}$ and $\theta_{\rm EW}$ respectively denote the QCD and electroweak 
origins for the CP phase of the quark mass term. 
Thus the statistical significance of $R_H$ will keep biggest 
unless the accuracy in measuring the Higgs mass and top Yukawa coupling gets better than the level of $\sim 10^{-32}$, the size of $R_H$.

The fine-tuning problem that we presently address is nothing sensory, but is essentially related to existence of a hidden symmetry which makes $R=1$ or $\chi_{\rm top}/m_l^2=0$ or $\chi_{\rm chiral} = \chi_{\rm axial}$ in terms of our estimator. 
This is in contrast to the conventional argument: 
Something delicately fine or not is controlled merely whether the tuning to some extent is necessary to make a big cancellation.
The quantification of the present fine-tuning is unambiguously made on the basis of the statistical significance, and the standard deviations are subject to the accuracy in the measurement of $R$s, which should therefore be compared with those having the same level of accuracy,  
as done above.

Note that the strong CP problem is trivially solved in the limit $m_l \to 0$, whereas the new fine-tuning problem gets more serious ($R \to 0$). 
This discrepancy in the two problems 
can be understood via the Ward identity Eq.(\ref{WI-def}), 
where $\chi_{\rm top}$ itself can be sent to zero when $m_l \to 0$, 
which is the massless up quark solution to the strong CP problem, 
however, $\chi_{\rm top}/m_l^2$ then blows up, leaving 
the new fine-tuning problem. 
Thus, the two problems are generically separated.
This fact also proves that the chiral $SU(2)$ symmetry (with $m_l \to 0$) 
does not make QCD free from the fine-tuning, 
in sharp contrast to the naive folklore.

We could start with the definition of $R$, $R= \chi_{\rm chiral}/\chi_{\rm axial}$, 
instead of the Ward identity Eq.(\ref{WI-def}), and discuss the difference $\Delta_{\rm axial-chiral} = \chi_{\rm axial} - \chi_{\rm chiral}$, so that $R = 1 - \Delta_{\rm axial-chiral}/\chi_{\rm axial}$.  Then the form of Eq.(\ref{WI-def}) is unambiguously fixed as it stands, 
and tells us that the symmetric limit $R=1$ is realized when $\chi_{\rm top}/m_l^2 \to 0$, which cannot be made when $m_l \to 0$ (because $\chi_{\rm top} \sim m_l$ for small $m_l$), but can be achieved when $m_s \to 0$, which is based on the symmetry argument. 
This alternative view would also help readers to more definitely see that the $m_l=0$ limit 
separates the new fine-tuning problem from the strong CP problem, where 
$R \to 0$ and $R_{\theta} = \theta_{\rm QCD}/\theta_{\rm EW} \to 1$, respectively.

\section{A candidate solution: new quarks with dark replica of QCD colors}

\textcolor{black}{
The proposed new fine-tuning problem 
is present at the scale only around the order of 1 GeV. 
When the electroweak symmetry becomes manifest 
at higher scales,  
the fine-tuning problem will be obscure 
because it explicitly breaks the global chiral $SU(2)$ and $U(1)$ axial symmetries as well as quark masses (or Yukawa couplings between the Higgs and quarks), so that the key Eq.(\ref{WI-def}) 
will be modified involving the electroweak ``topological" susceptibilities. 
Also at scales $\lesssim m_\pi$, 
the fine-tuning will be nontransparent due to 
the decoupling of pions, which is the most dominant source 
to generate the big gap between $\chi_{\rm chiral}$ and $\chi_{\rm axial}$. 
Thus the new fine-tuning problem needs to be solved by a new physics with the scale $\Lambda_{\rm NEW}$ in a range of $m_\pi \lesssim \Lambda_{\rm NEW} \lesssim$ 1 GeV. 
}


\textcolor{black}{
The hint for this avenue is seen in Eq.(\ref{tech-natur-def}), which indicates introducing massless new quarks. 
In fact, the topological susceptibility $\chi_{\rm top}$ 
goes to zero, when a massless new quark couples to the other three quarks, 
due to the flavor-singlet nature, so that Eq.(\ref{tech-natural-chi}) is realized. 
The detailed proof is given in Appendix~\ref{App:B}. 
This motivates one to consider an explicit model beyond the SM. 
}

We consider a new chiral quark ($\chi$) to be neutral under the electroweak charges, instead, carries a dark color of $SU(N_d)$ group under the fundamental representation.  The group representation table for the $\chi$ quark 
thus goes like $\chi_{L,R} \sim (N_d, 3, 1)_0$ 
for $SU(N_d) \times SU(3)_c \times SU(2)_W \times U(1)_Y$, where 
the latter three symmetries correspond to the SM's ones 
(QCD color, weak, and hypercharge). 
The dark color symmetry as well as the electroweak neutrality forbids creating undesired light hadrons composed of the ordinary light quarks and the $\chi$ quark, such as $\bar{u} \chi$ and $u u \chi$.

For simplicity, 
we also assume that the dark QCD coupling $g_d$ gets strong almost at the same scale 
as the ordinary QCD coupling does, 
$(\Lambda_d \sim \Lambda_{\rm QCD} 
= {\cal O}(1) \, {\rm GeV})$, namely,  
$g_d \sim g_s$.

Below the scale $\sim $ 1 GeV, 
the dark QCD dynamically breaks the dark chiral $U(3)_L \times U(3)_R$ symmetry for the $\chi$ quark, down to the vectorial part, where the extra factor of 3 in the number of flavors comes from the QCD color multiplicity. Only the hadrons singlet under 
both the dark and ordinary QCD colors survive in the vacuum. 
Then there emerges only one composite Nambu-Goldstone boson, 
$\eta_d$, which becomes pseudo 
due to the axial anomaly in the dark QCD sector and acquires the mass of ${\cal O}(1)$ GeV. 
Besides, at almost the same scale, ordinary QCD breaks the 
approximate chiral $SU(3 + N_d)_L \times SU(3 + N_d)_R$ symmetry involving the $\chi$ quark down to the vectorial one, where again only the color singlets are relevant. 
The spontaneous breaking of this extended chiral symmetry does not yield excessive meson spectra made of the ordinary quarks, because of the double color symmetries, as aforementioned. 
Thus the new low-lying spectra consist only of the dark sector: 
$\eta_d \sim \bar{\chi} i \gamma_5 \chi$ and its dark chiral 
partner $\sigma_d \sim \bar{\chi} \chi$, as well as 
spin-1 dark mesons ($\bar{\chi}\gamma_\mu \chi$ and $\bar{\chi} \gamma_\mu \gamma_5 \chi$), 
and dark baryons ($\sim \chi\chi\chi \cdots \chi$).  
All those low-lying dark hadrons have the mass on the order of 
1 GeV, by feeding the chiral breaking contributions from 
both the ordinary QCD and dark QCD sectors. 

\textcolor{black}{
Relaxing the new fine-tuning is tied with vanishing curvature of the QCD vacuum at around the QCD scale: $\chi_{\rm top}\to 0$. 
The new fine-tuning problem is present irrespective to the place of the QCD vacuum, i.e., the value of the (net) QCD $\bar{\theta}$ parameter: even a shifted 
QCD vacuum with $\bar{\theta}$ gone, say by assumption of a QCD axion, keeps nonzero  
$\chi_{\rm top}$ as the developed axion potential energy including 
the axion mass, which takes precisely the same flavor-singlet form as in Eq.(\ref{chitop-small-ml}). 
Thus the new fine-tuning problem is definitely separated from 
the strong CP problem.}

\textcolor{black}{
Actually, the dark QCD solution instead implies a nontrivial relation between $\bar{\theta}$ and $\theta_d$, the theta parameter in dark QCD, to realize $\chi_{\rm top} = 0$ in the presence of the massless new $\chi$ quark: 
the anomalous axial rotation of the massless $\chi$ 
leaves the $\bar{\theta}$-dependence into the dark QCD topological sector, $(\theta_d + (N_c/N_d) \bar{\theta}) Q^d_{\rm top}$, where 
$Q_{\rm top}^d$ denotes the dark QCD topological charge. 
Thereby, one gets $\chi_{\rm top} = (N_c/N_d)^2 \chi_{\rm top}^d \neq 0$  
with $\chi_{\rm top}^d$ being the topological susceptibility in dark QCD, unless $\theta_d = - (N_c/N_d) \bar{\theta}$. 
This nontrivial relation is required no matter what size  $\bar{\theta}$ is, i.e., which is independent of the strong CP problem. 
}

\textcolor{black}{
The required relation might be trivial when 
QCD itself relaxes $\bar{\theta}$ to 0 at an deep infrared fixed point to be consistent with realization of the confinement, 
as recently discussed in lattice QCD~\cite{Nakamura:2021meh,Nakamura:2019ind,Schierholz:2022wuc}. 
This self-relaxation is applicable also to dark QCD, hence 
in that case one has $\bar{\theta} = \theta_d =0$, and $\chi_{\rm top} = 0$ 
in the presence of the massless $\chi$ quark. 
}

\textcolor{black}{
In contrast, solving the new fine-tuning problem disfavors 
QCD axion models as the solution of the strong CP: 
first of all, the QCD axion needs not to be present 
until the QCD pions are decoupled from $R$, otherwise 
the axion potential energy necessarily yields nonzero 
$\chi_{\rm top}$ even along with massless new quarks. 
In this sense a composite axion~\cite{Kim:1984pt,Choi:1985cb} with the dynamical/composite scale scaled down to the QCD scale (or lower) might be the candidate, where the composite scale 
is set to $\sim m_\pi \sim 4 \pi f_a$ with the axion decay 
constant $f_a$. However, such a low-scale QCD axion model with both a QCD axion and its small decay constant on the QCD scale has already been 
ruled out by the LEP search for $Z \to \pi \gamma$~\cite{Jaeckel:2015jla} 
due to too large axion coupling to diphoton. 
Thus the QCD axion is incompatible with the solution for 
the new fine-tuning problem. 
}

In the next couple of subsections, 
we will discuss several characteristic features of the presently introduced dark QCD model 
below and above the scale $\sim 1$ GeV, in the aspects of phenomenology 
as well as astrophysical and cosmological observations.

\subsection{$\eta_d - \eta'$ mixing}

\begin{figure}[t] 
\includegraphics[width=6cm,clip]{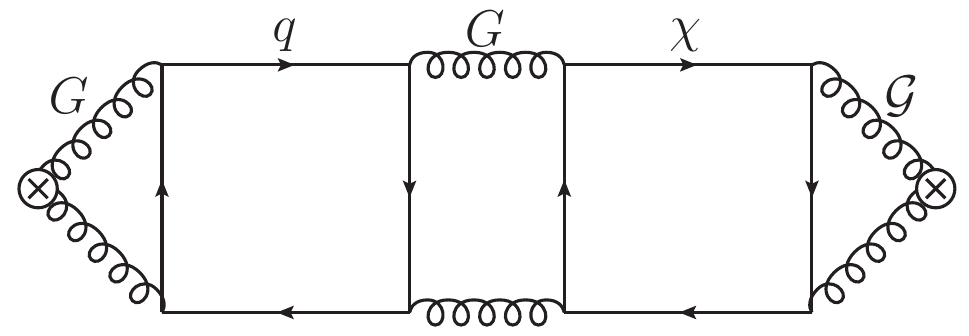}
\includegraphics[width=6cm,clip]{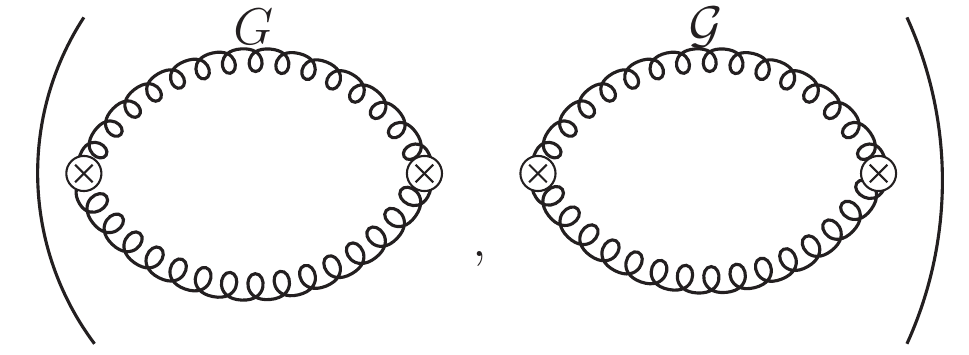}

\caption{
Left: A Feynman graph yielding the cross-term amplitude $\langle  
N_c ({\cal G}_{\mu\nu} \tilde{\cal G}^{\mu\nu}) \cdot  
N_d (G_{\mu\nu} \tilde{G}^{\mu\nu})  \rangle$. 
Right: the one for the non-mixing terms 
$\langle N_c ({\cal G}_{\mu\nu} \tilde{\cal G}^{\mu\nu}) \cdot  
N_c ({\cal G}_{\mu\nu} \tilde{\cal G}^{\mu\nu})   \rangle$ 
and 
$\langle N_d (G_{\mu\nu} \tilde{G}^{\mu\nu}) \cdot 
N_d (G_{\mu\nu} \tilde{G}^{\mu\nu})  \rangle$. 
Those diagrams correspond to the leading order in the large-$N_c$ and 
-$N_d$ expansion. 
The symbol $\otimes$ denotes the axial current source.  
Spring lines 
stand for propagators of QCD gluons ($G$) and dark QCD gluons (${\cal G}$), 
and straight lines for 
ordinary quarks ($q$) and new $\chi$ quark ($\chi$).  
}
\label{etad-etaprime}
\end{figure}

The $\eta_d$ dark meson can mix with the ordinary QCD $\eta'$ by sharing 
the axial anomaly via QCD. The mixing can be seen through 
the non-conservation law of the dark axial current $J^\mu_{\eta_d} = \bar{\chi} \gamma_\mu \gamma_5 \chi$: 
\begin{align} 
\partial_\mu J^\mu_{\eta_d} 
= 
N_d \frac{g_s^2}{32 \pi^2} (G_{\mu\nu} \tilde{G}^{\mu\nu}) 
+ N_c \frac{g_d^2}{32 \pi^2} ({\cal G}_{\mu\nu} \tilde{\cal G}^{\mu\nu}) 
\,,
\end{align}
with $N_c=3$ and $(G_{\mu\nu} \tilde{G}^{\mu\nu})$ and 
$({\cal G}_{\mu\nu} \tilde{\cal G}^{\mu\nu})$   
being the topological operators of the ordinary QCD and dark QCD, respectively. 
The size of mixing with $\eta'$, which couples to the $(G_{\mu\nu} \tilde{G}^{\mu\nu})$ term, 
can be evaluated by 
constructing the two-point correlation function of 
$(\partial_\mu J^\mu_{\eta_d})$, 
$\langle  (\partial_\mu J^\mu_{\eta_d}) 
(\partial_\nu J^\nu_{\eta_d}) 
$, 
 and focusing on the cross-term amplitude $\langle  
N_c ({\cal G}_{\mu\nu} \tilde{\cal G}^{\mu\nu}) \cdot 
N_d (G_{\mu\nu} \tilde{G}^{\mu\nu})   \rangle$. 
Estimate of the order of magnitude for the mixing amplitude can be made by working on 
the dual large-$N_c$ and -$N_d$ expansion with $\alpha_s \sim 1/N_c$ and $\alpha_d \sim 1/N_d$. 
Since the mixing amplitude arises necessarily due to the $\chi$ quark loop, 
it is estimated to be on the order of  ${\cal O} (N_c N_d \alpha_d \alpha_s \Lambda_d^2 \Lambda_{\rm QCD}^2) = {\cal O}(\Lambda_{\rm QCD}^2 \Lambda_{d}^2)$. 
For Feynman diagram interpretation, see Fig.~\ref{etad-etaprime}. 
This is compared with the non-mixing term amplitudes $
\langle N_c ({\cal G}_{\mu\nu} \tilde{\cal G}^{\mu\nu}) \cdot 
N_c ({\cal G}_{\mu\nu} \tilde{\cal G}^{\mu\nu}) \rangle 
= {\cal O}(\alpha_d N_c^2 N_d^2 \Lambda_d^4) = {\cal O}(N_c^2 N_d \Lambda_d^4) $, 
dominated by the dark gluon loop, and 
$\langle N_d (G_{\mu\nu} \tilde{G}^{\mu\nu}) \cdot  
N_d (G_{\mu\nu} \tilde{G}^{\mu\nu})  \rangle = {\cal O}(\alpha_s N_c^2 N_d^2 \Lambda_{\rm QCD}^4)={\cal O}(N_c N_d^2 \Lambda_{\rm QCD}^4)$, 
dominated by the ordinary QCD gluon loop. 
The size of the mixing angle $\theta_{\eta'-\eta_d}$ is then estimated as    

\begin{align} 
  |\theta_{\eta'-\eta_d}|  
  & \sim 
  \Bigg| 
  \frac{N_c ({\cal G}_{\mu\nu} \tilde{\cal G}^{\mu\nu}) \cdot  
  N_d (G_{\mu\nu} \tilde{G}^{\mu\nu})  }{\langle 
  N_d (G_{\mu\nu} \tilde{G}^{\mu\nu}) \cdot 
  N_d (G_{\mu\nu} \tilde{G}^{\mu\nu})   \rangle 
  - 
  \langle N_c ({\cal G}_{\mu\nu} \tilde{\cal G}^{\mu\nu}) 
  \cdot 
  N_c ({\cal G}_{\mu\nu} \tilde{\cal G}^{\mu\nu})   \rangle} 
  \Bigg| 
  \notag\\ 
 & 
  = 
  {\cal O}\left(  
  \Bigg| 
  \frac{\alpha_d \alpha_s \Lambda_d^2 \Lambda_{\rm QCD}^2}{ N_c N_d (\alpha_s  \Lambda_{\rm QCD}^4 - \alpha_d \Lambda_d^4)}
  \Bigg|
  \right) 
  = {\cal O}\left( \frac{\alpha_s}{N_c N_d}  \right) 
  =
   {\cal O}(10^{-3}) \times \left( \frac{\alpha_s}{0.3} \right) \left( \frac{5}{N_d} \right) 
\,, 
\end{align}

\noindent 
where in the last second equality we have used $\alpha_s \sim \alpha_d$ and $\Lambda_{\rm QCD} \sim \Lambda_{d}$ in magnitude, and in the last one a typical size of $\alpha_s$ at around 1 GeV~\cite{Deur:2016tte}, 
$\alpha_s (1\,{\rm GeV}) \simeq 0.3$, 
has been quoted as a reference value, which is precisely the same 
order of magnitude as expected from the large $N_c$ scaling (i.e., 
$\alpha_s \sim 1/N_c \sim$ 30\%).  
The mixing angle is of ${\cal O}(10^{-2})$ in unit of degree, hence   
can safely be neglected compared to the mixing between 
$\eta'$ and $\eta$ in the ordinary QCD, which is estimated to be 
about $28^\circ$ from the recent lattice simulation at 
physical point for 2 +1 flavors~\cite{Bali:2021qem}, 
consistently with the prediction from the chiral perturbation theory, $\sim 20^\circ$~\cite{Gasser:1984gg} 
and also the current Particle Data Group. 
Thus the successful $\eta'$ physics in ordinary QCD is not substantially altered. 

\subsection{Probing dark mesons via couplings to photon}

\begin{figure}[t] 
\includegraphics[width=7cm,clip]{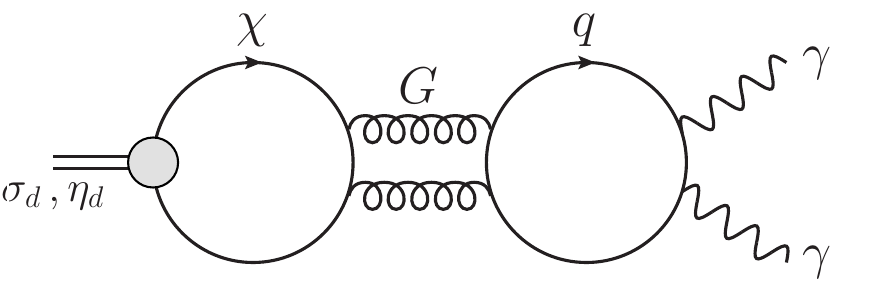}

\caption{
A typical Feynman graph describing 
generation of 
$\sigma_d$ - $\gamma$ - $\gamma$ and $\eta_d$ - $\gamma$ - $\gamma$ 
vertices. 
Graphical notations are the same as those in Fig.~\ref{etad-eta}  
}
\label{etad-2gamma}
\end{figure}

$\eta_d \sim \bar{\chi} i \gamma_5 \chi$ and $\sigma_d \sim \bar{\chi} \chi$  
can couple to the ordinary quarks through the $\chi$ loop with two gluon exchanges. 
Therefore, they can also couple to diphoton through the ordinary quark loops 
with the loop-induced $\eta_d$-digluon and $\sigma_d$-digluon vertices. 
See Fig.~\ref{etad-2gamma}. 
The large-$N_c$ and -$N_d$ counting can evaluate 
the order of magnitude for couplings to diphoton at the nontrivial leading order.
The induced photon couplings can be summarized by the following effective Lagrangian: 
\begin{align} 
& {\cal L}_{ a \gamma\gamma} 
=  g_{\eta_d \gamma\gamma} \frac{\eta_d}{4} 
F_{\mu\nu} \tilde{F}^{\mu\nu} 
+ 
  g_{\sigma_d \gamma\gamma} \frac{ \sigma_d}{4} 
F_{\mu\nu} {F}^{\mu\nu}  
\,, \notag\\ 
& g_{\eta_d(\sigma_d)  \gamma\gamma} 
 \sim 
\frac{\alpha_{\rm em} \alpha_s^2}{ \pi ( 4\pi)^2 f_d} N_c^2 N_d \sum_{q} [Q_{\rm em}^q]^2 
\,, 
\end{align}
where $\alpha_{\rm em}$ is the fine structure constant for 
the electromagnetic coupling, which is $\simeq 1/137$ at the scale of order of 1 GeV, 
and $Q_{\rm em}^q$ denotes the electromagnetic charge for the ordinary quark $q$ in the 
standard model. 
The dark-meson decay constant $f_d$ can be related to the intrinsic scale $\Lambda_d$ as 
$f_d \sim  \sqrt{N_d}/(4\pi) \cdot \Lambda_d$ with 
the large $N_d$ scaling taken into account. 
Then the size of the coupling to diphoton 
can be estimated as 
\begin{align} 
g_{\eta_d(\sigma_d)  \gamma\gamma} 
={\cal O}(10^{-4}) \, {\rm GeV}^{-1} 
\times 
\left(\frac{N_d}{5} \right) \left( \frac{\rm GeV}{\Lambda_d}  \right) 
\,. 
\end{align}

$\eta_d$ and $\sigma_d$ with this size of photon coupling at the mass around 1 GeV  
are too short-lived (with the lifetime $\tau \sim 10^{-12}$ s) 
to have sensitivity for astrophysical observations, such as 
cosmic ray telescopes, but can be probed by collider experiments 
in the same manner as the axionlike particle (a) searches. 
The currently available observation limit around the target mass and photon coupling 
comes from  the Belle II experiment at the SuperKEKB collider on $e^- e^+ \to 3 \gamma$ event 
with $496\,{\rm pb}^{-1}$ data. 
This experiment has placed the upper bound on the photon coupling,  
$g_{a \gamma \gamma} \lesssim 10^{-3}\,{\rm GeV}^{-1}$~\cite{Belle-II:2020jti}. 
The Belle II future prospects with $20\,{\rm fb}^{-1}$ and $50\,{\rm ab}^{-1}$ data 
will reach $g_{a \gamma \gamma} \sim 10^{-4}\,{\rm GeV}^{-1}$ at the mass around 1 GeV~\cite{Dolan:2017osp}, 
which can probe $\eta_d$ and $\sigma_d$ via the $3 \gamma$ signal. 
More precise estimate on the $g_{\eta_d (\sigma_d) \gamma \gamma}$ coupling 
is necessary to give more definite prediction to the $3 \gamma$ event, which 
will be pursed elsewhere.

\begin{figure}[t] 
\includegraphics[width=7cm,clip]{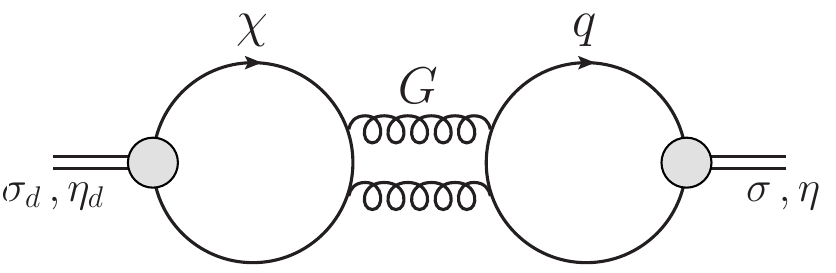}

\caption{
A Feynman graph generating 
the $\sigma_d$ - $\sigma$ and $\eta_d$ - $\eta$ mixings. 
The filled blob implies nonperturbative formation of those bound states 
with the strong (Yukawa-type) coupling of ${\cal O}(1)$. 
Other graphical notations are the same as those in Fig.~\ref{etad-etaprime}.  
}
\label{etad-eta}
\end{figure}

 \subsection{More on dark meson - ordinary meson mixing} 

Through loop processes generating the $\eta_d$ - and $\sigma_d$ - photon couplings 
with external photon legs replaced by the ordinary QCD mesons, 
$\eta_d$ and $\sigma_d$ can also mix with the ordinary mesons in the isosinglet 
channel, such as $\eta$ and $\sigma = f_0(500)$. See Fig,~\ref{etad-eta}. 
This goes like quadratic-mass mixing, the order of which can be 
evaluated by the large -$N_c$ and -$N_d$ expansion, 
to be $N_d \alpha_s^2 \Lambda_d^2/(4\pi)^4 = 
[{\cal O}(1) \, {\rm MeV} \times 
\left(\sqrt{\frac{N_d}{5}} \right) \left( \frac{\alpha_s}{0.3} \right) \left( \frac{\Lambda_d}{\rm GeV}
 \right)]^2$. Referred to the diagonal masses $m_{\eta_d} \sim m_{\sigma_d} 
 = {\cal O}(1)$ GeV and $m_\sigma \sim m_\eta \sim 500$ MeV, 
 the size of the mixing angle is estimated as $\frac{1}{2} |\frac{{\rm MeV}^2}{500\,{\rm MeV}^2}| \, {\rm radian} \sim 0.03^\circ$. 
 This mixing is, again, safely negligible in comparison 
 with the $\eta-\eta'$ mixing (by the angle $\sim 20^\circ -30^\circ$), and can be insensitive to 
 highly uncertain and large mixing structure of the isosinglet mesons in QCD. 
A similar argument is applicable also to other mesons with higher spins, like vector and 
axialvector mesons. Thus the successful hadron physics is intact.




\subsection{Cosmological abundance of dark baryon}

The dark baryons are formed as color-singlet for both ordinary QCD and 
dark QCD colors. The wavefunction of the ground-state spin-1/2 dark baryon takes the form, e.g., for $N_d=5$, like 
$n_d \, \sim \epsilon^{C' D' E'} Q_{C'} Q_{D'} Q_{E'} $ 
with 
$ Q_{C'} = \epsilon_{ABC}(\chi^A_{i_1} \chi^B_{i_2} \chi^C_{i_3}) \otimes 
\epsilon_{A'B'C'}(\chi^{A'}_{i_4}\chi^{B'}_{i_5})f^{i_1 i_2 i_3 i_4 i_5}$, 
where the symbols $\epsilon$'s and $f$ denote the group structure constants for 
$SU(3)_c$ and $SU(N_d=5)$ groups, respectively. 
The dark baryon mass $m_{n_d}$ scales as $\sim N_d \cdot N_c$ in the large-$N_c$ and -$N_d$ expansion, 
so that it can be larger than the masses of the lightest dark mesons $\eta_d$ and $\sigma_d$, 
to be $m_{n_d} \sim (N_c N_d/3) m_n$ normalized to the ordinary QCD baryon mass $m_n \sim$  1 GeV.

The dark baryon is completely stable due to the exact dark baryon number conservation in dark QCD, 
hence can be a dark matter. 
Since the $\chi$ quark can be thermalized with the ordinary QCD gluons 
in the thermal plasma of the standard model, 
the dark baryon $n_d$ as well as the dark mesons $(\sigma_d, \eta_d, \cdots)$ 
could be thermally produced as well. 
As in the case of the ordinary baryons, 
$n_d$ could annihilate into the lightest meson pairs, i.e., via $n_d \bar{n}_d \to \eta_d \eta_d$, 
or more multiple $\eta_d$ states (and also into $\sigma_d$ states), 
which would determine the freeze out of the number density of $n_d$.

Rough, but, conservative estimate of the thermal relic abundance of $n_d$ 
could be done by assuming the standard freeze-out scenario for this annihilation 
with the size of the classical cross section $\langle \sigma v \rangle \sim 4 \pi/m_{n_d}^2$. 
This crude approximation could be justified in the  
large-$N_c$ and $-N_d$ limit, where the dark baryon behaves like almost 
static, nonrelativistic, and a classical rigid body with finite 
radius (i.e. impact parameter) of ${\cal O}(1/m_{n_d})$. 
The thermal relic abundance is evaluated as~\cite{KT} 
$\Omega_{n_d} h^2 \sim \frac{10^9 x_{\rm FO}}{\sqrt{g_*(T_{\rm FO})} M_{\rm p} {\rm GeV} \langle \sigma v \rangle}$, where $x_{\rm FO} = m_{n_d}/T_{\rm FO}$ with $T_{\rm FO}$ being the 
freeze-out temperature; $M_{\rm pl}$ is the Planck scale $\sim 10^{18}$ GeV; 
$g_*(T_{\rm FO})$ the effective degree of freedom of relativistic particle at $T=T_{\rm FO}$. 
The standard freeze-out scenario gives $x_{\rm FO}\sim 20$, and $g_*(T_{\rm FO}) = {\cal O}(50)$ at $T$ below $1$ GeV~\cite{Zyla:2020zbs}. 
The relic abundance of $n_d$ is then estimated to be of ${\cal O}(10^{-9})$ for 
$m_{n_d} = {\cal O}(N_d=3-5)$ GeV, which 
is compared with the observed abundance of the cold dark matter (CDM) today $\sim 0.1$. 
 Thus the dark baryon $n_d$ cannot fully 
 account for the presently observed dark matter abundance.  
 The model needs to be improved to be completed, so as to include 
 another dark matter, which yields the main component of the CDM abundance 
 today. 

\subsection{Direct detection of dark baryon}

Regarding the direct detection experiments by recoils  
of heavy nuclei with dark matters, 
the dark matter $n_d$ with mass of ${\cal O}(N_d)$ GeV could contribute to the spin-independent  
scattering cross section. 
$n_d$ strongly couples to $\eta_d$ and $\sigma_d$, 
which can convert into the ordinary $\eta$ and $\sigma$ mesons through 
the mixing with angle $\theta_{\sigma-\sigma_d}$ of ${\cal O}(10^{-3})$ in unit of radian,  
as discussed above. 
Since the pseudoscalar-mediator contribution vanishes at the leading order 
at zero momentum 
transfer relevant to such a nonrelativistic scattering process, 
the dominant contribution would come from the $\sigma_d$ - $\sigma$ 
meson-mixture portal process. 
Analogously to the Higgs portal scenario, 
the spin-independent dark matter - nucleon ($N$) cross section $\sigma_{\rm SI}(n_d N \to n_d N)$ 
can then be evaluated as 
\begin{align} 
 & \sigma_{\rm SI}(n_d N \to n_d N) 
 \notag\\ 
 & 
 \sim \frac{\theta_{\sigma-\sigma_d}^2 m_*^2(m_N, m_{n_d})}{16 \pi | \sqrt{s_{\sigma}^{\rm pole}} |^4} g_{\sigma_d n_d n_d}^2 g_{\sigma NN}^2   
\,, \label{sigma-SI}
\end{align}
where $m_*(m_N, m_{n_d}) \equiv m_N m_{n_d}/(m_N + m_{n_d})$; $\sqrt{s^{\rm pole}_{\sigma}}$ 
denotes the complex pole mass of $\sigma$ meson ($f_0(500)$)  
in the $T$-matrix measured at experiments; 
$g_{\sigma_d n_d n_d}$ is the $\sigma_d$ coupling to the dark baryon $n_d$, 
$g_{\sigma_d n_d n_d} \sim \frac{m_{n_d}}{f_d} \sim \frac{4 \pi}{\sqrt{N_d}} \frac{m_{n_d}}{\Lambda_d}$; 
$g_{\sigma N N}$ stands for the $\sigma$ meson coupling to nucleon, which can be 
evaluated via so-called the nucleon $\sigma$ term $\sigma_{\pi N} = \langle N| m_l \bar{q}_l q_l |N \rangle$ as 
$g_{\sigma NN} = \frac{\sigma_{\pi N}}{f_\pi}$.

We assume that the $n_d$-dark matter has the same velocity distribution in the dark matter halo as in the case of CDM.  
Using $\sigma_{\pi N} \simeq 0.05m_N$~\cite{Ohki:2008ff,Ohki:2012jyg,Hisano:2015rsa}, 
$f_\pi\simeq 0.0924$ GeV, $m_N\simeq 0.940$ GeV, and $\sqrt{s^{\rm pole}_\sigma} \sim (0.5 -  0.3 i)$ GeV~\cite{Zyla:2020zbs}, 
we then have $\sigma_{\rm SI}(n_d N \to n_d N) \times \frac{\Omega_{n_D} h^2}{\Omega_{\rm CDM} h^2} 
\sim 10^{-43}\, {\rm cm}^2$ 
at $m_{n_d}=5$ GeV, 
for $N_d=5$, $m_{\sigma_d}=1$ GeV, $\Lambda_d = 1$ GeV, 
$\theta_{\sigma-\sigma_d}=10^{-3}$, and $\Omega_{n_d} h^2=10^{-9}$. 
This signal can be explored by the planning detection experiment 
searching for sub-GeV dark matters, called ALETHEIA~\cite{Liao:2021npo}.

Incorporation of the $\eta_d$-portal contribution at loop-level might be crucial at 
this order of the cross section, and might enhance the cross section as discussed 
in the literature~\cite{Abe:2018emu}. 
Furthermore, the gluonic-nucleon matrix element $\langle N| \frac{\alpha_s}{\pi} G_{\mu\nu}^2  |N \rangle$ together with the operator $\bar{n}_d n_d G_{\mu\nu}^2  $ 
could also contribute to the cross section, which might be comparable with the $\sigma_d-\sigma$ portal 
contribution in Eq.(\ref{sigma-SI}). 
More detailed analysis is to be performed elsewhere.

%


\subsection{Collider detection prospect of dark hadrons}

Exotic hadrons can be created 
as hybrid bound sates of the ordinary QCD quarks and 
the dark $\chi$ quarks. 
Those are forced to form like a molecular type, such as 
$\bar{q}_i q_j \bar{\chi}\chi$ and 
$\bar{q}_iq_j \chi \chi \cdots \chi$, because 
of the QCD and dark QCD color symmetries. 
The lightest ones would be four-quark 
bound states made of 
the ordinary QCD pions ($\pi$) and $\eta_d$ or $\sigma_d$. 

In a view of collider phenomenology, the four-quark states $(\pi \eta_d)$ and $(\pi \sigma_d)$
finally decay to 
$4 \gamma$ or $\mu(e)$ + $2 \gamma$ + missing energy with a few GeV scale.  
Those exotic hadrons could be produced 
at hadron colliders through rescattering of $\pi$ and $\eta_d$ or $\sigma_d$, or $n_d$, in which   
the dark QCD hadrons could be produced via the gluon fusion process, 
while the ordinary QCD pion via the initial-gluonic bremsstrahlung almost 
colinear to the produced dark hadrons. 
The signal events as above 
would be hard to detect over the huge backgrounds with hadronic jets with energy on the GeV scale,   
hence is so challenging at the current status of hadron collider experiments. 
A photon-photon collider with high accuracy in photon production events with the GeV scale might be more practical to explore those signals.

\subsection{Constraints from QCD running coupling}

Extra massless quarks contribute to the running evolution of 
the ordinary QCD coupling $g_s$, where 
in the present case 
$N_d$ species of new quarks in the fundamental representation of $SU(3)_c$ group 
come into play. 
To keep the asymptotic freedom, at least $N_d$ has to be $< (33/2) -6 \sim 10$, 
which is determined by the one-loop perturbative calculation. 
Collider experiments have confirmed the asymptotic freedom with 
high accuracy 
in a wide range of higher energy scales, in particular above 10 GeV, over 1 TeV~\cite{Zyla:2020zbs}. 
When $\alpha_s$ is evolved up to higher scales 
using $\alpha_s(M_Z)$
measured at the $Z$ boson pole 
as input,  
the tail of the asymptotic freedom around ${\cal O}(1)$ TeV 
can thus have sensitivity to exclude new quarks.

Current data on $\alpha_s$ at the scale around ${\cal O}(1)$ TeV  
involve large theoretical uncertainties.  
This results in uncertainty of determination of $\alpha_s(M_Z)$ for 
various experiments (LHC-ATLAS, -CMS, and Tevatron-CDF, and D0, etc.), 
which yields $\alpha_s(M_Z) \simeq 0.110 - 0.130$, being consistent with 
the world average $\alpha_s(M_Z)\simeq 0.118$ 
within the uncertainties~\cite{Zyla:2020zbs}.

We work on the two-loop perturbative computation of $\alpha_s$. 
The dark QCD running coupling ($\alpha_d$) contributes to the running of $\alpha_s$ at two-loop level.  
This contribution is, however, safely negligible: 
when $\alpha_s \sim \alpha_d$ at low-energy 
scale as desired, because 
$\alpha_d \ll \alpha_s$ at high energy due to 
the smaller number of dark QCD quarks (with the net number 3 coming in the beta function 
of $\alpha_d$) 
than that of the ordinary QCD quarks (with the net number 5 or 6 + $N_d$ in the beta function 
of $\alpha_s$).

Taking into account 
only the additional $N_d$ quark loop contributions to the running of $\alpha_s$, 
we thus compute the two-loop beta function, and find that 
as long as $N_d$ is moderately large $(N_d \le 5)$, 
the measured ultraviolet scaling (for 
the renormalization scale of $\mu$ = 10 GeV $-$ a few TeV) 
can be consistent with the current data~\cite{Zyla:2020zbs} within 
the range of $\alpha_s(M_z)$ above.

Precise measurements in lower scales $\lesssim 10$ GeV 
have not well been explored so far, 
due to the deep infrared complexity of QCD. 
The low-energy running of $\alpha_s$ is indeed still uncertain, and can be variant as 
discussed in a recent review, e.g., \cite{Deur:2016tte}. 
The present dark QCD could dramatically alter the infrared running feature of $\alpha_s$, 
due to new quarks and the running of the dark QCD coupling $\alpha_d$. 
This will also supply a decisive answer to a possibility of 
the infrared-near conformality of the real-life QCD, to which 
point we will come back in the conclusion section.

{\color{black}
Other stringent bounds on the extra light quarks or colored scalars 
come from the ALEPH search for gluino and squark pairs tagged with the multijets 
at Large Electron Positron (LEP) collider experiment~\cite{ALEPH:1997mcm}. 
However, this limit has no sensitivity below the mass $\sim 2$ GeV, 
hence is not applicable to the present benchmark model. 
}

Thus a few massless new quarks can still survive 
constraints on $\alpha_s$ at the current status. 
More precise measurements of $\alpha_s$ in the future will clarify  
how many light or massless new quarks can be hidden in QCD, 
which will fix the value of $N_d$ in terms of the present dark QCD.



\section{Other theoretical remarks on fine-tuning free QCD} 

 Finally, we 
argue 
 one nontrivial dynamical issue related to successful realization of the fine-tuning free condition 
 with massless new quarks.

\subsection{A nontrivial dynamical issue}

It is true that $\chi_{\rm top}  = 0$ and $\chi_{\eta}=\chi_\pi$ when 
massless new quarks are assumed to be present, but 
one might naively think that it implies $m_\pi \sim m_\eta$ 
because the size of susceptibility 
follows the associated meson mass, as noted in Sec.~III. 
This would obviously be contradicted with the observation. 
However, it should not be the case.

First of all, 
the susceptibilities correspond to meson-correlation functions 
at zero momentum transfer, at off-shell of mesons, 
so do not exactly equal to propagators of mesons. 
Pions are light enough, and supposed to have the mass close to 
such soft momenta, and $\chi_\pi$ gets affected only from 
the up and down quark loops [see Eq.(\ref{chipi}) with 
Eqs.(\ref{pion}) and (\ref{chipi-NJL}) in Appendix~\ref{App:A}]. 
Therefore, $\chi_\pi$ would almost scale with $1/m_\pi^2$. 
In contrast, the $\eta$ meson mass is thought to be far off soft, and indeed 
$\chi_\eta$ gets contributions not only from $u$ and $d$ quark 
loops, but also the strange quark's [see Eq.(\ref{chi-eta}) with Eq.(\ref{chip}) in Appendix~\ref{App:A}], 
where the latter contribution is crucial 
to yield $\chi_\eta < \chi_\pi$ and does not simply follow 
the inverse mass scaling $\sim 1/m_\eta^2$.

Now consider the case with the new $\chi$ quark, where  
the loop corrections to $\chi_\eta$ involve the $\chi$ quark term, 
as well as the other three quarks terms.  
[$\chi_\eta$ will be constructed from the pseudoscalar susceptibilities 
labelled as $\chi_P^{00}, \chi_P^{08}, \chi_P^{015}, \chi_P^{88}, \chi_P^{815}$, and $\chi_P^{1515}$ associated 
with the generators of $U(4)$.] 
The $\chi$ loop corrections would thus be a key  
to realize $\chi_\eta = \chi_\pi$, while keeping $m_\eta > m_\pi$,  
which would make the model parameter space of the dark QCD limited 
(e.g. the separation of QCD and dark QCD gauge couplings in size would be constrained).

More precise discussion is subject to explicit nonperturbative 
computation of susceptibilities as well as meson spectra with the $\chi$ quark in QCD which also couples to dark QCD. 
This could be possible by lattice simulations, or also 
by working on NJL-like models, or applying the 
chiral perturbation theory. 
This is, however, 
beyond the current scope, to be left as the important dynamical issue, 
in the future.

\section{Conclusion}

Toward the deeper understanding of the flavor-dependence of the QCD vacuum, 
in the present work 
we have focused on a gap between the breaking strengths of the chiral $SU(2)$ symmetry for the light up and down 
quarks and $U(1)$ axial symmetry. 
It might be too premature to conclude that this gap is 
related merely to the mystery of the quark-generation structure in the SM. 
We have found an alternative interpretation based on the symmetry argument: 
the gap can be relaxed by a chiral symmetry for the strange quark, and the role of the massless/light strange quark can be replaced by a new massless/light quark (called $\chi$).

The existence of the symmetric limit $m_s\to 0$, where the ``chiral $SU(2)$ = $U(1)$ axial" is realized as above, is manifest in the flavor dependence of the QCD vacuum, and we have shown how the gap, i.e., the fine-tuning is seriously large, on the basis of the statistical analysis along with comparison with the existing fine-tuned quantities.

Thus, QCD may be yet incomplete and 
QCD of the SM calls for 
more quarks to keep the equivalence of the strengths of the 
chiral $SU(2)$ and $U(1)$ axial breaking, to be free from the fine-tuning. 
New quarks need protected to be (nearly) massless by a new chiral symmetry, and can be introduced in QCD consistently with existing experiments, as demonstrated in the present work.   
This symmetry is independent of the existing chiral or isospin symmetry,  which ensures the smallness of masses of  
light quarks and had so far played the role to make  
QCD free from the fine-tuning, e.g., for the small proton - neutron mass 
difference compared to individual proton and neutron masses.

The new fine-tuning problem is triggered   
due to the large strange quark mass, and brings 
a big gap between the chiral and axial order 
parameters, detected as a small size of the estimator $R$ in Eq.(\ref{R-def}). 
This trend actually persists even in a whole temperature, as can be 
understood by tracing the analysis in the recent literature~\cite{Cui:2021bqf}. 
This small $R$ can be checked on the lattice QCD simulations in the future.

In this paper, a benchmark of the fine-tuned QCD was placed based on 
the well-known low-energy effective model, NJL.
This is a pioneer step, and should be motivated 
and confirmed by various approaches in the future, 
such as the lattice calculation and functional-renormalization group analysis. 
One can also work on the chiral perturbation theory 
to evaluate $R$. No work has so far been done properly taking into account the flavor singlet condition for 
$\chi_{\rm top}$, hence it would also be an interesting issue to be left in the future.

Besides the phenomenological and cosmological consequences discussed in Sec.~VI,   
the notion of the fine-tuning free QCD potentially provides rich perspective toward  
deeper understanding of the real-life QCD today and past in the 
thermal history of the universe. 
Several of those are listed below.  

\begin{enumerate}

\item[(i)] 

In the thermal history of the universe, 
massless new quarks should have contributed to the thermal QCD phase transition.  
As was clarified by the nonperturbative renormalization group analysis~\cite{Braun:2010qs} 
and the lattice simulation~\cite{Lombardo:2014mda}, 
a large number of light quarks generically decrease the critical temperature of 
the chiral symmetry restoration, $T_c$; e.g., $T_c$ gets smaller by a factor of 
about 2/3 and 1/3~\cite{Lombardo:2014mda}, when the number of quarks is increased from 3 to 6 (corresponding to $N_d=3$ 
in the dark QCD scenario) and 8 ($N_d=5$), respectively. 
The lattice QCD simulations with 2 +1 flavors at physical point 
have reported the pseudo-critical temperature of the chiral 
crossover, $T_{\rm pc} \simeq 155$ MeV~\cite{Aoki:2006we,Borsanyi:2011bn,Ding:2015ona,HotQCD:2018pds,Ding:2020rtq}. When this $T_{\rm pc}$ is simply scaled by the above flavor dependence, 
we have $T_{\rm pc} \simeq 103$ MeV for $N_d=3$, and $T_{\rm pc}\simeq 52$ MeV for $N_d=5$. 
In particular, this reduction of $T_{\rm pc}$ 
would give a significant impact on evaluation of 
the cosmological abundance of dark matters related via 
the scattering with, or the annihilation into, quarks, when 
the freeze-out temperature is in a range from ${\cal O}(T_{\rm pc}/10)$ to 
${\cal O}(10 T_{\rm pc})$. 
Even in this refinement, 
the dark baryon dark matter would still yield a negligibly small thermal abundance. 
When a popular Higgs-portal dark matter, which is sensitive to the Higgs invisible decay as well, 
is for instance additionally considered, 
this refinement of $T_{\rm pc}$ would suggest 
substantial re-analysis on a viable parameter space.

\item[(ii)]

When the massless new quarks have a new color like ``$N_d$" of dark QCD in the benchmark model 
introduced in the present paper, 
the scenario of the chiral phase transition will be complicated and rich, where 
the net chiral phase transitions will be categorized into two : 
the dark-chiral phase transition related to the 3 massless quarks will be 
first order, which will take place at around the conventional $T_c$, meanwhile  
the chiral phase transition (crossover) in QCD associated with the three light $3 + N_d$ quarks 
would follow at around $T=T_c/3 - 2T_c/3$, for $N_d= 5 -3$. 
The latter crossover might undergo a ``jump" at around $T_c$ due to 
the interference with dark QCD colored quarks which undergoes the first order phase transition. 
This would generate a sharp drop at around $T=T_c$ characteristic to the first-order phase transition, 
to deform the conventional chiral crossover to be like a first-order type. 
Exploration of the feasibility of such an induced ``jump" 
would be worth pursing by lattice simulations, or 
the nonperturbative renormalization group analysis, or 
chiral effective models such as the NJL model. 
The nonperturbative analysis on an NJL model including QCD as well as dark QCD interactions, which is sort of an extended variant of so-called the gauged NJL model~\cite{Kondo:1988qd,Appelquist:1988fm,Kondo:1991yk,Kondo:1992sq,Kondo:1993ty,Kondo:1993jq,Harada:1994wy,Kubota:1999jf},   
would also be intriguing to pursue.

\item[(iii)]

If the chiral crossover experiences a ``jump" in the QCD phase transition epoch 
of the thermal history of the universe as argued in (ii), 
nonzero latent heat might also be promptly created during the first-order like 
phase transition. 
The associated bubbles might then be nucleated to be expanded and collided each other over the Hubble evolution, and develop gravitational waves, to reach us today, to be detected by gravitational wave interferometers~\cite{Kosowsky:1992rz,Kosowsky:1992vn,Caprini:2015zlo}. 
Thus, the evidence of the chiral phase transition in QCD, 
namely the origin of nucleon mass in the thermal history, might directly be probed by gravitational wave interferometers in the future. 
This would be an innovative probe of the first-order QCD phase transition 
without invoking a flavor-dependent phase transition at 
the vicinity of the critical endpoint expected to be present with nonzero 
baryon chemical potential. 
The dedicated study on the gravitational wave signal 
would be necessary, to be left in a separated publication.

\item[(iv)]

If the number of massless (light) new quarks is five, as in the case of 
one of the presently addressed dark-QCD benchmark models, 
the real-life QCD of $SU(3)$ group would possess eight flavors at the low-energy. 
This implies that actually, the real-life QCD might be what is called walking dynamics, 
having the infrared-near conformality characterized by the Caswell-Banks-Zaks infrared fixed point~\cite{Caswell:1974gg,Banks:1981nn}.  
Therefore,  
it would be worth investigating the fine-tuning free QCD in depth, 
even in the context of such an infrared conformality of QCD. 
Actually, the infrared feature would be more intricate, because the 
 five new  quarks make the dark QCD communicated with the ordinary QCD in the renormalization group equations. The Caswell-Bank-Zaks infrared fixed point 
 would thus be generalized in the two-coupling space $(\alpha_s, \alpha_d)$, which 
 is particularly noteworthy to explore. 

\item[(v)] 
The fine-tuning free QCD might share somewhat similarity with 
three-flavor conformal QCD in the  literature~\cite{Crewther:2013vea,Crewther:2020tgd}, 
or a walking gauge theory with many fermion flavors extensively discussed 
in a view of the dynamical-electroweak symmetry-breaking scenario~\cite{Yamawaki:1985zg,Bando:1986bg,Bando:1987we,Akiba:1985rr,Appelquist:1986an,Appelquist:1987fc,Holdom:1984sk}. 
Clarification of the style of the low-energy effective theory 
for the fine-tuning free QCD would thus be of interest. 
It might be nontrivial, in particular, 
whether the real-life QCD could reflect the almost scale-invariant nature 
in the chiral Lagrangian or  not~\cite{Matsuzaki:2013eva,Crewther:2013vea,Li:2016uzn,Kasai:2016ifi,Hansen:2016fri,Appelquist:2017wcg,Appelquist:2017vyy,Cata:2019edh,Appelquist:2019lgk,Brown:2019ipr,Golterman:2020utm}.

\item[(vi)]

The walking QCD indicated in (iv) and (v), as the fine-tuning free QCD, would also be relevant to a longstanding question on whether the $\sigma$ meson in QCD 
could be a composite pseudo Nambu-Goldstone boson, so-called the pseudo dilaton, associated with the spontaneous-scale symmetry breaking in QCD. 
The lattice simulations on eight flavor QCD with fundamental representation fermions 
have proven the evidence of the light flavor-singlet 
$\bar{q}q$ meson, identified as the composite 
dilaton, 
with mass comparable with the mass of the pseudo Nambu-Goldstone boson (like pion in the conventional QCD) associated with the spontaneous breaking 
of the eight-flavor chiral symmetry~\cite{LatKMI:2014xoh}. 
To properly match the currently favored walking QCD setup, 
it would be necessary to work on QCD with 2 + 1 + 5 flavors, where the latter five light fermions 
are charged also under dark QCD of $SU(5)$ group, and measure the lightest flavor-singlet 
$\bar{q}q$ meson signal, which is identified as the $\sigma$ meson in the real-life QCD, and then 
confirm the $\sigma$ meson mass consistent with the experimentally observed value. 
This study might lead to a heuristic solution to the aforementioned question on the QCD dilaton, 
and would simultaneously resolve complexity of the scalar meson puzzle with 
the striking answer of no significant mixing with four-quark states, nor glueballs.


\end{enumerate}

In closing, 
other than the dark QCD model addressed 
in the present paper, it would be worth investigating 
modeling beyond the SM 
with taking into account making QCD fine-tuning free. 

Here is the recipe: 
first of all, one needs to introduce new Dirac massless quarks, which  
act as a spectator of the global chiral $SU(2)$ symmetry for 
the up and down quarks. 
They are generically allowed to feel the electroweak charge, 
whichever way it is ganged vectorlikely or chirally. 
The former case would be phenomenologically viable in light of 
the electroweak precision tests, and the limit on the number 
of quark generations placed from the $Z$ boson decays. 
Second, those new quarks would be preferable not to form the 
Yukawa interaction with ordinary quarks and Higgs fields 
(doublets and triplets, and so on) which 
develop the vacuum expectation values at the weak scale, 
and yield the mass for the new quarks. 
If new quarks could be coupled to such Higgses, solving to the strong CP problem 
(without introducing an axion) 
as well as keeping the light enough new quarks down until the QCD scale 
would be hard and challenging.

Given this recipe, one might think that though 
it would sound somewhat ad hoc, 
the presumably most minimal setup would be to introduce a  electroweak-singlet quark 
with a negative charge under a new parity, 
while assign a positive charge for ordinary quarks, 
so as to avoid spoiling the successful light hadron spectroscopy. 
In the dark QCD model introduced in the present paper, 
the role of such a new parity has been played by the dark QCD color charge. 
Such alternatives are to be addressed in details elsewhere.




\section*{Acknowledgements} 


We thank Ryosuke Sato for useful comments and discussions. 
This work was supported in part by the National Science Foundation of China (NSFC) under Grant No.11975108, 12047569, 12147217 
and the Seeds Funding of Jilin University (S.M.).  
The work of A.T. was supported by the RIKEN Special Postdoctoral Researcher program
and partially by JSPS  KAKENHI Grant Number JP20K14479.

\appendix

\section{Scalar and pseudoscalar susceptibilities}\label{App:A}

 In this Appendix,  the explicit formulae for scalar and pseudoscalar susceptibilities derived from the NJL model Eq.(\ref{Lag:NJL}) are listed. 
 We will leave the details of the calculations here and refer readers to a review paper~\cite{Hatsuda:1994pi}, which contains all necessary information to reach the final formulas that we will present below. 
 In what follows, we use the notation for quark condensates as 
 $\alpha = \langle \bar{u}u \rangle$, 
  $\beta = \langle \bar{d}d \rangle$, 
  and 
   $\gamma = \langle \bar{s}s \rangle$, and will work on computations 
   in Euclidean momentum space.

\subsection{Pseudoscalar meson channel}

In the $
\eta$ - $\eta \prime$ coupled channel, the pseudoscalar meson susceptibility on  the generator basis of $U(3)$ is defined as 
\begin{equation}
	\chi_P^{ij} = \int d^4 x \langle (i \bar q (x) \gamma_5 \lambda^i  q (x))(i \bar q (0)\gamma_5 \lambda^j  q(0)) \rangle\,, 
\label{chiP-ij}
\end{equation}
 where $i,j=0,8$.  
In the NJL model, after prescribing  
the resummation technique, this $\chi_P^{ij}$ takes the form~\cite{Hatsuda:1994pi}  
\begin{equation} 
	\chi_{P}=\frac{-1}{1+{\bf G}_{P}\Pi_{P}} \cdot \Pi_{P} 
\,, \label{chiP-formula}
\end{equation}
where ${\bf G}_P$ is the coupling strength matrix and $\Pi_P$ is the polarization tensor evaluated at zero momentum transfer: 
\begin{align}
{\bf G}_P 
&=
\begin{pmatrix}
	{\bf G}_P^{00} & {\bf G}_P^{08}\\
	{\bf G}_P^{80} & {\bf G}_P^{88}
\end{pmatrix}
\notag\\ 
&=
\begin{pmatrix}
	G_S-\frac{2}{3}(\alpha +\beta+\gamma )G_D &  -\frac{\sqrt{2}}{6} (2\gamma-\alpha-\beta )G_D \\
	-\frac{\sqrt{2}}{6} (2\gamma-\alpha-\beta )G_D & G_S-\frac{1}{3}(\gamma-2\alpha -2\beta )G_D
\end{pmatrix}
\,, \label{Gs-p}
\end{align} 
\begin{equation}
\Pi_P=
\begin{pmatrix}
	\Pi _P^{00} & \Pi _P^{08}\\
	\Pi _P^{80} & \Pi _P^{88}
\end{pmatrix}
=
\begin{pmatrix}
	\frac{2}{3}	(2I_P^{uu}+I_P^{ss}) &  \frac{2\sqrt{2}}{3}(I_P^{uu}-I_P^{ss}) \\\frac{2\sqrt{2}}{3}(I_P^{uu}-I_P^{ss} ) & \frac{2}{3} (I_P^{uu} + 2I_P^{ss})
\end{pmatrix}
\,, 
\end{equation}
with $I_P^{ii}(\omega, \boldsymbol{p})$ being the pesudoscalar one-loop polarization functions, 
and the relevant quantities: 
\begin{align} 
	I_P^{ii} &=-\frac{N_c}{\pi ^2}\int^\Lambda _0  d p\, p^2  \frac{1}{E_{i}} 
	\,, \qquad {\rm for } \quad i=u,d,s 
\,, \notag\\ 
E_i &= \sqrt{M_i^2 + p^2} 
\,, \notag\\ 
M_u &=  m_l-2G_S \alpha - 2G_D\beta \gamma 
\,, \notag\\ 
M_d &= m_l -2G_S \beta - 2G_D \alpha \gamma 
\,, \notag \\
M_s&=m_s-2G_S \gamma - 2G_D\alpha \beta
\,, \notag\\ 
	\langle  \bar q_i q_i \rangle 
	& 
	= - 2N_c\int^{\Lambda}\frac{d \boldsymbol{p} }{(2\pi)^3} \frac{M_i}{E_{i}} 
	\,. \label{IPij}
\end{align}

By performing the basis transformation, 
the pseudoscalar susceptibilities in the flavor basis are obtained as  
\begin{equation}
	\begin{pmatrix}
	\frac{1}{2}\chi_P^{uu}
	+\frac{1}{2}\chi_P^{ud}
	\\
	\chi_P^{us} \\
	\chi_P^{ss}
	\end{pmatrix}
=
	\begin{pmatrix}
	\frac{1}{6} & \frac{\sqrt2}{6} & \frac{1}{12} \\
    \frac{1}{6} & -\frac{\sqrt2}{12} & -\frac{1}{6} \\
    \frac{1}{6} & -\frac{\sqrt2}{3} & \frac{1}{3}
	\end{pmatrix}
	\begin{pmatrix}
	\chi_P^{00} \\
	\chi_P^{08} \\
	\chi_P^{88}
	\end{pmatrix}
	\label{chip}
	\,, 
\end{equation}
where we have taken the isospin symmetric limit into account, i.e., $\chi_P^{uu} = \chi_P^{dd}$ and $\chi_P^{us}=\chi_P^{ds}$. 
The $\chi_\eta$ is then given as in Eq.(\ref{chi-eta}).


In a way similar to Eq.(\ref{chiP-formula}), 
the present NJL model gives the explicit formula of $\chi_\pi$ defined in Eq.(\ref{psesus}) as 
\begin{equation}
	\chi_{\pi}=\frac{-1}{1+{\bf G}_{\pi}\Pi_{\pi}} \cdot \Pi_{\pi}
	\label{pion}
\,, 
\end{equation}
where ${\bf G}_{\pi} = G_S + G_D \gamma$, which is the coupling strength in the pion channel, and $\Pi_{\pi}$ is the quark-loop polarization function for $\chi_{\pi}$, which is evaluated 
by using $I_P^{ii}$ in Eq.(\ref{IPij}) as  
\begin{equation}
	\Pi_{\pi}=I_P^{uu}+I_P^{dd}=2I_P^{uu}
\,. \label{chipi-NJL}
\end{equation}

\subsection{Scalar meson channel}

The definitions of scalar susceptibilities are similar to those for pseudoscalars', which are given just by removing {\it{$i\gamma_5$}} in the definition of pseudoscalar susceptibilities, 
and supplying  
the appropriate one-loop polarization functions and the corresponding coupling constants.

In the same way as in the pseudoscalar susceptibilities in the 
0 - 8 coupled channel in Eq.(\ref{chiP-ij}), 
the scalar susceptibility matrix $\chi_S$ is evaluated in the 
present NJL on the generator basis as 
\begin{equation}
	\chi_{S}=\frac{-1}{1+{\bf G}_{S}\Pi_{S}} \cdot \Pi_{S} 
\,, \label{chi-S}
\end{equation}
where ${\bf G}_{S}$ is the coupling strength matrix, 
\begin{align}
	{\bf G}_S 
	&=
\begin{pmatrix}
	{\bf G}_S^{00} & {\bf G}_S^{08}\\
	{\bf G}_S^{80} & {\bf G}_S^{88}
\end{pmatrix}
\notag\\ 
& =
\begin{pmatrix}
	G_S+\frac{2}{3}(\alpha +\beta+\gamma )G_D &  \frac{\sqrt{2}}{6} (2\gamma-\alpha-\beta )G_D \\
	\frac{\sqrt{2}}{6} (2\gamma-\alpha-\beta )G_D & G_S+\frac{1}{3}(\gamma-2\alpha -2\beta )G_D
\end{pmatrix}
\,. \label{Gs-s}
\end{align}
The coupling constant matrices in the scalar and pseudoscalar channels (Eqs.(\ref{Gs-p}) and (\ref{Gs-s})) are different in sign in front of $G_D$, reflecting  
attractive and repulsive interactions, respectively.  
The scalar polarization tensor matrix $\Pi_S$ in Eq.(\ref{chi-S}) is given by  
\begin{equation}
\Pi_S=
\begin{pmatrix}
	\Pi _S^{00} & \Pi _S^{08}\\
	\Pi _S^{80} & \Pi _S^{88}
\end{pmatrix}
=
\begin{pmatrix}
	\frac{2}{3}	(2I_S^{uu}+I_S^{ss}) &  \frac{2\sqrt{2}}{3}(I_S^{uu}-I_S^{ss}) \\\frac{2\sqrt{2}}{3}(I_S^{uu}-I_S^{ss} ) & \frac{2}{3} (I_S^{uu} + 2I_S^{ss})
\end{pmatrix}
\,, 
\end{equation}
\begin{equation}
	I_S^{ii} =-\frac{N_c}{\pi ^2}\int^\Lambda _0 p^2 dp \frac{E_{ip}^2-M_i^2}{E_i^3} 
	\,, \qquad 
	{\rm for} \qquad i=u,d,s
\,. 
\end{equation}

By moving on to the flavor base via the base transformation, 
the scalar susceptibilities are cast into the form:
\begin{equation}
	\begin{pmatrix}
	\frac{1}{2}\chi_S^{uu}
	+\frac{1}{2}\chi_S^{ud}
	 \\
	\chi_S^{us} \\
	\chi_S^{ss}
	\end{pmatrix}
=
	\begin{pmatrix}
	\frac{1}{6} & \frac{\sqrt2}{6} & \frac{1}{12} \\
    \frac{1}{6} & -\frac{\sqrt2}{12} & -\frac{1}{6} \\
    \frac{1}{6} & -\frac{\sqrt2}{3} & \frac{1}{3}
	\end{pmatrix}
	\begin{pmatrix}
	\chi_S^{00} \\
	\chi_S^{08} \\
	\chi_S^{88}
	\end{pmatrix}
\,, 
\end{equation}
in which we have read $\chi_S^{uu}=\chi_S^{dd}$
and $\chi_S^{us}=\chi_S^{ds}$.


 Similarly to $\chi_\pi$ in Eq.(\ref{pion}), 
in the NJL model the explicit formula for $\chi_\delta$  
reads 
\begin{equation} 
	\chi_{\delta}=\frac{-\Pi_{\delta} }{1+{\bf G}_{\delta}\Pi_{\delta} }
	\label{delta}
	\,, 
\end{equation}
where ${\bf G}_{\delta} = G_S - G_D \gamma$, which is the coupling strength in the $\delta$ channel, and $\Pi_{\delta}=I_S^{uu}+I_S^{dd}=2I_S^{uu}$ is the corresponding quark-loop polarization function.

\section{Anomalous chiral-Ward identities and topological susceptibility, with one extra quark}\label{App:B}

This Appendix provides derivation of anomalous chiral 
Ward identities including one extra quark in three flavor QCD.

The anomalous chiral Ward identities are directly read off  
from chiral variations of the generating functional of 
the QCD action with $N$ quark flavors. 
The central formula then takes the form 
\begin{align} 
 \langle \delta_a {\cal O}_b(0) \rangle
 = i \int d^4 x \langle 
  {\cal O}_b(0) \cdot \bar{Q}(x) i \gamma_5 \{ T_a, M \} Q(x)   
 \rangle 
 \,, 
\end{align} 
where $Q$ denotes a quark field forming $N$-plet of $SU(N)$; 
$T_a$ ($a=1, \cdots, N^2-1$) are generators of $SU(N)$; 
$\delta_a$ stands for the infinitesimal variation of 
the chiral $SU(N)$ transformation associated with the generator $T_a$, under which $Q$ transforms as 
$\delta_a Q = i \gamma_5 T_a Q$; ${\cal O}_b(0)$ 
($b=0, \cdots, N^2-1$) is an arbitrary 
operator; 
$M$ is the $Q$-quark mass matric, taken to be diagonal, like 
$M= {\rm diag}\{m_1, m_2, \cdots, m_N \}$.

In particular, for the pseudoscalar operators 
${\cal O}_b = \bar{Q} i\gamma_5 T_b Q$, 
taking the case with $N=3$, with $Q= (q_l, s)^T= ((u,d),s)^T$, $m_1=m_2=m_l$, 
and $m_3=m_s$, and choosing $a=1,2,3, 8$ and $b=0, 8$, 
we get the first and second identities in Eq.(\ref{chipi}) in the main text.

Including a new quark field $\chi$ with mass $m_4 = m_\chi$ 
into the $Q$-field to form the $SU(4)$-quartet: 
$Q = (q_l, s, \chi)^T$, we work on the chiral $SU(4)$ transformations.
Among the $SU(4)$ generators,  we focus only on the subgroup part of $SU(3)$ 
embedded as $T_{a=1,\cdots, 8} = 
\frac{1}{2} 
\left( 
\begin{array}{cc}
\lambda_a & 0_{1 \times 3} \\
0_{3 \times 1} & 0 
\end{array} 
\right)$, and the Cartan generator $T_{a=15} = 
\frac{1}{2\sqrt{6}}  
\left( 
\begin{array}{cc}
- 1_{3 \times 3} & 0_{1 \times 3} \\
0_{3 \times 1} & 3 
\end{array} 
\right)
$, 
together with the unit matrix $T_{a=0} = \frac{1}{2\sqrt{2}} \cdot 1_{4 \times 4}$.  We thus find a couple of additional Ward identities, except for 
the first two in Eq.(\ref{chipi}): 

\begin{align} 
(a, b)=(8,0): \quad 
& 
\langle \bar uu \rangle +\langle \bar dd \rangle
- 2 \langle \bar{s}s \rangle \notag\\
&= - \left[ m_l 
\left(\chi_{P}^{uu}+\chi_{P}^{dd}+2\chi_{P}^{ud}\right)
+ (m_l - 2 m_s) 
\left(\chi_{P}^{us}+\chi_{P}^{ds} \right)
- 2 m_s \chi_P^{ss} 
+  m_\chi  
\left(\chi_{P}^{u\chi}+\chi_{P}^{d\chi} \right)
-2  m_s \chi_P^{s \chi} \right] 
\,; \notag\\ 
(a, b)=(15,8): \quad 
& 
\langle \bar uu \rangle +\langle \bar dd \rangle
- 2 \langle \bar{s}s \rangle
\notag\\
&= - \left[ m_l 
\left(\chi_{P}^{uu}+\chi_{P}^{dd}+2\chi_{P}^{ud}\right)
+ (m_s - 2 m_l)
\left(\chi_{P}^{us}+\chi_{P}^{ds} \right)
- 2 m_s \chi_P^{ss} 
- 3 m_\chi  
\left(\chi_{P}^{u\chi}+\chi_{P}^{d\chi} \right)
+ 6 m_\chi \chi_P^{s \chi} \right] 
\,; \notag\\ 
(a, b)=(8,15): \quad 
& 
\langle \bar uu \rangle +\langle \bar dd \rangle
- 2 \langle \bar{s}s \rangle 
\notag\\
&= - \left[ m_l 
\left(\chi_{P}^{uu}+\chi_{P}^{dd}+2\chi_{P}^{ud}\right)
+ (m_l - 2 m_s) 
\left(\chi_{P}^{us}+\chi_{P}^{ds} \right)
- 2 m_s \chi_P^{ss} 
- 3 m_l  
\left(\chi_{P}^{u\chi}+\chi_{P}^{d\chi} \right)
+ 6 m_s \chi_P^{s \chi} \right] 
\,; \notag\\ 
(a, b)=(15,0): \quad 
& 
\langle \bar uu \rangle +\langle \bar dd \rangle
+ \langle \bar{s}s \rangle 
- 3 \langle \bar{\chi} \chi \rangle 
\notag\\ 
& = - \Bigl[ m_l 
\left(\chi_{P}^{uu}+\chi_{P}^{dd}+2\chi_{P}^{ud}\right)
+ (m_l + m_s) 
\left(\chi_{P}^{us}+\chi_{P}^{ds} \right)
+ m_s \chi_P^{ss} 
+ (m_l - 3 m_\chi)  
\left(\chi_{P}^{u\chi}+\chi_{P}^{d\chi} \right)
\notag\\
&\qquad\;\;
+ (m_s - 3 m_\chi) \chi_P^{s \chi} 
- 3 m_\chi \chi_P^{\chi\chi}  \Bigl] 
\,; \notag\\ 
(a, b)=(15,15): \quad 
& 
\langle \bar uu \rangle +\langle \bar dd \rangle
+ \langle \bar{s}s \rangle 
+ 9 \langle \bar{\chi} \chi \rangle 
\notag\\ 
& = - \Bigl[ m_l 
\left(\chi_{P}^{uu}+\chi_{P}^{dd}+2\chi_{P}^{ud}\right)
+ (m_l + m_s) 
\left(\chi_{P}^{us}+\chi_{P}^{ds} \right)
+ m_s \chi_P^{ss} 
- 3 (m_l + m_\chi)  
\left(\chi_{P}^{u\chi}+\chi_{P}^{d\chi} \right)
\notag\\
&
- 3 (m_s +m_\chi) \chi_P^{s \chi} 
+ 9 m_\chi \chi_P^{\chi\chi}  \Bigl] 
\,, \label{WIs-SU4}
\end{align}
\noindent 
where 
pseudoscalar susceptibilities including the $\chi$ quark 
are defined in the same way as those for other quarks in Eq.(\ref{psesus}).
From these with Eq.(\ref{chipi}), 
we have 
\begin{align}
\left(\chi_{P}^{u\chi}+\chi_{P}^{d\chi} \right)
& = \frac{m_s}{m_\chi} 
\left(\chi_{P}^{us}+\chi_{P}^{ds} \right)
\,, 
\notag\\ 
\chi_P^{s \chi} 
& = \frac{m_l}{2 m_\chi} 
\left(\chi_{P}^{us}+\chi_{P}^{ds} \right)
\,, 
\notag\\ 
\langle \bar{\chi} \chi \rangle 
&= - m_\chi \chi_P^{\chi\chi} 
+ \frac{m_l m_s}{2 m_\chi} 
\left(\chi_{P}^{us}+\chi_{P}^{ds} \right)
\,. \label{chiP-chi}
\end{align}
Using the first two relations in Eq.(\ref{chiP-chi}), we see that 
the first Ward identity in Eq.(\ref{WIs-SU4}) becomes the same as the third one in Eq.(\ref{chipi}).

Including the new $\chi$ quark, 
the topological susceptibility defined in Eq.(\ref{chitop:def}) 
now takes the form 
\begin{align} 
\chi_{\rm top} 
&= 
 \bar{m}^2 \Bigg[ 
		 \frac{\langle \bar{u}u\rangle}{m_l}
		 +
		 \frac{\langle \bar{d}d\rangle}{m_l}
		 +\frac{\langle \bar {s}s \rangle}{m_s} 
		+ \frac{\langle \bar {\chi}\chi \rangle}{m_\chi} 
\notag\\ 
&		
        +  
		\left(\chi_{P}^{uu}+\chi_{P}^{dd}+2\chi_{P}^{ud}\right)
		+ 
	    2  
	    \left(\chi_{P}^{us}+\chi_{P}^{ds} \right)
	    +\chi_P^{ss} 
	    \notag\\ 
	    &
	    + 
		2  
		\left(\chi_{P}^{u\chi}+\chi_{P}^{d\chi} \right)
		+ 2 \chi_P^{s\chi}  +\chi_P^{\chi\chi}
        \Bigg]\,, 
\end{align} 
with $\bar{m}^{-1} = \left( \frac{2}{m_l} + \frac{1}{m_s} + \frac{1}{m_\chi} \right)$. 
Using the relations in Eq.(\ref{chiP-chi}) together with 
those in Eq.(\ref{chipi}), 
we find 
\begin{align} 
\chi_{\rm top} 
&= \frac{1}{2} m_l m_s 
\left(\chi_{P}^{us}+\chi_{P}^{ds} \right)
= \frac{1}{2} m_l m_\chi 
\left(\chi_{P}^{u\chi}+\chi_{P}^{d\chi} \right)
 =  m_s m_\chi \chi_P^{s \chi} 
\notag\\ 
& = 
\frac{1}{4} \left[ 
m_l\left(\langle \bar{u} u \rangle +\langle \bar{d} d \rangle  \right)
	+ m_l^2\left(\chi_{P}^{uu}+\chi_{P}^{dd}
	+2\chi_{P}^{ud}\right)
\right]
= m_s \langle \bar{s}s \rangle + m_s^2 \chi_P^{ss}
= m_\chi \langle \bar{\chi}\chi \rangle + m_\chi^2 \chi_P^{\chi\chi}
 \,. \label{chitop-chi} 
\end{align} 

In the case of dark QCD modeling as in the text, 
actually, the dark QCD coupling to the $\chi$ quark explicitly 
breaks the chiral $SU(4)_L \times SU(4)_R$ symmetry, as well as 
the mass terms. 
However, this breaking effect does not modify the anomalous 
identities associated with the chiral transformations for $a=1,2,3,8,15$, 
because the dark QCD coupling to the $\chi$ quark  only breaks the vectorial $SU(4)$ flavor symmetry 
down to $SU(3) \times U(1)$, which still keeps those chiral symmetries: 
$\left[T_a, \left( \begin{array}{cc} 
0_{3\times 3} & \\ 
 & 1 
\end{array}
\right) \right] \sim [T_a, T_{b=15}] =0$.

The argument in this Appendix 
can straightforwardly be extended to the case with more extra quarks, like $3 +N_d$ quarks.

\textcolor{black}{
The presence of the dark QCD theta parameter as well as 
the QCD's one modifies $\chi_{\rm top}$ in Eq.(\ref{chitop-chi}) as $\chi_{\rm top} \to \chi_{\rm top} + (N_c \bar{m}/m_\chi) \chi_{\rm top}^d$ with the dark QCD topological susceptibility $\chi_{\rm top}^d$, $N_c=3$, 
and $m_\chi$ in the original $\chi_{\rm top}$ as well as $ \bar{m}$ replaced by $m_\chi/N_d$. 
}



\end{document}